\documentclass[12pt,draftclsnofoot,onecolumn]{IEEEtran}
\usepackage{subfigure}
\usepackage[T1]{fontenc}
\usepackage{caption}
\renewcommand{\figurename}{Fig.}
\usepackage{algpseudocode}
\usepackage{algorithmicx,algorithm}
\usepackage{bm}
\usepackage{array}
\usepackage{booktabs}
\usepackage{mathtools}
\usepackage[font={small}]{caption}
\def\BibTeX{{\rm B\kern-.05em{\sc i\kern-.025em b}\kern-.08em
		T\kern-.1667em\lower.7ex\hbox{E}\kern-.125emX}}

\renewcommand{\baselinestretch}{1.46}

\usepackage{cite}

\ifCLASSINFOpdf
   \usepackage[pdftex]{graphicx}
   \graphicspath{{../pdf/}{../jpeg/}}
   \DeclareGraphicsExtensions{.pdf,.jpeg,.png}
\else
   \usepackage[dvips]{graphicx}
   \graphicspath{{../eps/}}
   \DeclareGraphicsExtensions{.eps}
\fi
\usepackage{amsmath}
\usepackage[cmintegrals]{newtxmath}
\begin{document}

\title{Environment-Aware Hybrid Beamforming by Leveraging Channel Knowledge Map}

\author{
	\IEEEauthorblockN{Di Wu,~\IEEEmembership{Student Member, IEEE}, Yong Zeng,~\IEEEmembership{Member, IEEE}, Shi Jin,~\IEEEmembership{Senior Member, IEEE} and Rui Zhang,~\IEEEmembership{Fellow, IEEE}}

	\thanks{
		D. Wu, Y. Zeng, and S. Jin are with the National Mobile Communications Research Laboratory, Southeast University, Nanjing 210096, China. Y. Zeng is also with the Purple Mountain Laboratories, Nanjing 211111, China (e-mail: \{230228195, yong\_zeng, jinshi\}@seu.edu.cn). (\emph{Corresponding author: Yong Zeng.})
		
		R. Zhang is with the Department of Electrical and Computer Engineering, National University of Singapore, Singapore 117583 (e-mail:\{elezhang@nus.edu.sg\}).
		
		This work was supported by the National Key R\&D Program of China with Grant number 2019YFB1803400.
		Part of this work has been presented in IEEE ICC 2021, Montreal, Canada, 14-23 June 2021 \cite{wu2021environment}.
	}
}

\maketitle

\begin{abstract}
Hybrid analog/digital beamforming is a promising technique to realize millimeter wave (mmWave) massive multiple-input multiple-output (MIMO) systems cost-effectively. 
However, existing hybrid beamforming designs mainly rely on real-time channel training or beam sweeping to find the desired beams, which incurs prohibitive overhead due to a large number of antennas at both the transmitter and receiver with only limited radio frequency (RF) chains. 
To resolve this challenging issue, in this paper, we propose a new  \emph{environment-aware} hybrid beamforming technique that requires only light real-time training, by leveraging the useful tool of channel knowledge map (CKM) with the user's location information.
CKM is a site-specific database, which offers location-specific channel-relevant information to facilitate or even obviate the acquisition of real-time channel state information (CSI). 
Two specific types of CKM are proposed in this paper for hybrid beamforming design in mmWave massive MIMO systems, namely \emph{channel angle map} (CAM) and \emph{beam index map} (BIM). 
It is shown that compared with existing environment-unaware schemes, the proposed environment-aware hybrid beamforming scheme based on CKM can drastically improve the effective communication rate, even under moderate user location errors, thanks to its great saving of the prohibitive real-time training overhead.
\end{abstract}


%
\IEEEpeerreviewmaketitle

\section{Introduction}
Millimeter wave (mmWave) massive multiple-input multiple-output (MIMO) systems are expected to achieve enormous communication rates, by jointly exploiting the ultra-wide bandwidth available at mmWave spectrum and the extremely high spectrum efficiency offered by massive MIMO communication \cite{MIMO,overview,rappaport2013millimeter}.
However, compared to the conventional sub-6GHz communication systems, mmWave signals suffer from more severe free-space propagation loss, which requires the base station (BS) and user equipment (UE) to be equipped with more antennas for compensating the propagation loss.
As such, the resultant high hardware cost, power consumption, and signal processing complexity of mmWave massive MIMO systems make it practically challenging to implement the conventional fully digital baseband processing such as transmitter/receiver precoding/combining, which requires each antenna element be connected to a costly radio frequency (RF) chain.
To address this issue, various cost-effective beamforming techniques have been proposed, by either reducing the cost per RF chain or using fewer RF chains than the number of antennas. 
For example, by using low-resolution analog-to-digital converters (ADC) (e.g., 1-3 bits per sample), the circuit power consumption can be significantly reduced \cite{alkhateeb2014mimo}. 
On the other hand, analog beamforming with all array elements sharing one single RF chain is another effective approach to reduce the cost of RF chains \cite{venkateswaran2010analog}.  
To further increase the communication rate by enabling spatial multiplexing, hybrid analog/digital beamforming has been extensively studied, which divides the signal processing operations into the analog and digital domains, and can be realized by either analog phase shifters \cite{zhang2005variable,Spatially} or lens antenna arrays \cite{zeng2016millimeter}. 
However, to practically reap the promised gains by hybrid beamforming, the acquisition of accurate channel state information (CSI) is essential, which is practically difficult since the system can only access the signals in much lower dimension than the number of antennas due to the limited RF chains. 

There are two major approaches in the literature to practically realize mmWave hybrid beamforming, namely {\it training-based CSI estimation} \cite{lee2016channel,venugopal2017channel,alkhateeb2014channel} and {\it training-based beam sweeping} \cite{xiao2017channel,wu2017non,zhou2021high}. 
In the first approach, the full MIMO CSI matrix is firstly estimated by pilot-based channel training, based on which the transmit and receive beamforming vectors are designed.
However, different from fully digital MIMO systems, channel estimation for mmWave massive MIMO with hybrid beamforming generally requires more substantial training overhead\cite{howmu}. 
This is because not only an excessively large number of channel coefficients need to be estimated, but also the channel measured in digital baseband is intertwined with the analog beamforming used in the RF domain to train the channel, and thus the actual MIMO channel matrix cannot be directly obtained.
While techniques such as compressive sensing (CS) can be applied to resolve this issue to certain extent \cite{cs}, they require high computational complexity to design the observation matrix and sophisticated iterative algorithms for signal processing. 
Besides, large training overhead is still required to obtain a high observation gain.
On the other hand, for the training-based beam sweeping approach\cite{tutora,heng2021six,kim2014elements,kim2014fast}, the transmit and receive beams are selected from predefined codebooks via sequential beam sweeping, without having to estimate the MIMO channel explicitly.
However, as the number of antennas or beam codebook size increases, this approach still incurs prohibitive training overhead to search over all possible combinations of transmit and receive beam pairs.
For example, as pointed out in \cite{tutora}, even for analog beamforming alone, with 64 antennas at the BS and 16 antennas at the UE, it may take up to 5.2 seconds (s) for sweeping over all the possible beams, which is unaffordable for practical systems.

It is worth mentioning that all the aforementioned techniques for mmWave beamforming mainly rely on real-time channel training to find the desired beams, while ignoring the user's location and its  actual communication environment. 
However, with the rapid advancement of RF-based localization and sensing technologies, utilizing the UE location information \cite{rezaie2020location,satyanarayana2019deep,va2019online,xiao2022overview} and even the geolocation-based database \cite{R1, R2, R3,va2017inverse,CKM,wu2021environment,ding2021environment,li2021channel} in wireless communication systems has received fast-growing attention recently. 
In particular, in \cite{CKM}, a novel concept of {\it channel knowledge map} (CKM) was proposed,
which aims to enable environment-aware communications by offering location-specific (rather than the conventional coarse site-specific) information regarding the intrinsic radio channels, so as to facilitate or even avoid the sophisticated real-time CSI acquisition.
Therefore, compared to the prevalent environment-unaware communications that only rely on real-time channel training, CKM-based environment-aware communication is expected to achieve significant performance gains, especially for scenarios when real-time channel training is costly or even impossible. 
Four typical such scenarios are outlined in \cite{CKM}, namely channels for yet-to-reach locations, channels for non-cooperative nodes, channels with large dimensions and channels under severe hardware/processing limitations.
In the preliminary version of this work \cite{wu2021environment}, CKM is utilized to design {\it environment-aware and training-free} analog beam alignment for mmWave massive MIMO systems, where the transmitter and receiver each has a single RF chain and can support one data stream only. 


In this paper, we extend our previous study on CKM-enabled environment-aware analog beam alignment in \cite{wu2021environment} to mmWave massive MIMO communication with the general hybrid analog/digital beamforming.
The main contributions of this work are summarized as follows:
\begin{itemize}
\item First, we present the codebook-based hybrid analog/digital beamforming architecture for mmWave massive MIMO communication, where the analog beams are selected from predetermined codebooks and the digital beams can be freely designed. 
To provide the performance upper bound for the considered system, we first assume that perfect CSI is available and derive the optimal analog beam selection and digital beamforming design schemes to achieve the maximum communication rate.
It is revealed that such a CSI-based optimal hybrid beamforming scheme not only incurs extremely large training overhead to acquire the CSI, but also needs prohibitive computational complexity to find the best beams for transmitter and receiver.
\item To resolve the above issues, we propose the environment-aware hybrid beamforming schemes, by leveraging the useful tool of CKM and the UE's location information, which is attainable in contemporary wireless systems. 
Two specific types of CKM are proposed, namely the {\it channel angle map} (CAM) and {\it beam index map} (BIM).
Specifically, CAM aims to predict the location-specific channel angle information, i.e., angle of arrivals (AoAs) and angle of departures (AoDs) of the potential channel paths associated with each possible UE location. 
A CAM-enabled light-training scheme is thus proposed to reconstruct the MIMO channel matrix, which greatly reduces the training overhead, thanks to the location-specific angle information provided by CAM. 
In contrast, BIM aims to provide the location-specific candidate transmit and receive beams for all possible UE locations.
While CAM still needs further computation for the optimal beams after estimating the explicit MIMO channel matrix, BIM-enabled scheme directly obtains beamforming matrices based on the reduced-dimension effective channel via light training, thus avoiding the prohibitive computational  complexity.
\item Last, extensive simulations are performed based on communication environment generated from the commercial ray-tracing software Wireless Insite\footnote{https://www.remcom.com/wireless-insite-em-propagation-software}.
The simulation results demonstrate that compared to various benchmark schemes such as CS-based and location-based methods, the proposed CKM-enabled environment-aware hybrid beamforming schemes can significantly improve the effective communication rate, even under moderate UE location errors, thanks to their environment-awareness for significantly saving the real-time training overhead. 
\end{itemize}

The organization for the rest of  this paper is as follows. 
Section II presents the system model for codebook-based mmWave massive MIMO systems.
In Section III, under the assumption of perfect CSI, the optimal scheme for analog beam selection and digital beamforming design is proposed, which, however, requires prohibitive training overhead and computational complexity. 
In Section IV, we propose the environment-aware hybrid beamforming method enabled by CAM and BIM, which can significantly reduce the real-time training overhead and/or computational complexity. 
In Section V, extensive numerical results are provided based on channels generated by commercial ray-tracing software, which demonstrate the advantages of the proposed schemes over various benchmark methods in terms of effective communication rate and training overhead. 
Finally, conclusion is drawn in Section VI.

We use the following notations throughout this paper.
Scalars are denoted by italic letters.
Boldface lower- and upper-case letters denote vectors and matrices, respectively. 
$ \mathbb{C}^{M\times N} $ and $ \mathbb{R}^{M\times N} $ denote the space of $ M \times N $ complex and real matrices, respectively.
$ \mathbf{A}^{T},\mathbf{A}^{*},\mathbf{A}^H $ denote the transpose, conjugate and conjugate transpose of the matrix $ \mathbf{A} $, respectively.
$ ||\mathbf{A}||_F $, $ \mathrm{tr}(\mathbf{A}) $, and $ |\mathbf{A}| $ denote the Frobenius norm, trace, and determinant of $\mathbf{A}$, respectively.
$ \mathbf{A}\succeq 0 $ means that $ \mathbf{A} $ is a positive semi-definite matrix. 
$ [\mathbf{A}]_{{\mathcal{R},:}} $ denotes a submatrix of $ \mathbf{A} $ formed by  rows with indices in the set $ \mathcal{R} $. 
Similarly, $ [\mathbf{A}]_{:,\mathcal{R}} $ is the submatrix formed by the columns with indices in $ \mathcal{R} $.
$ \mathrm{Diag}(\mathbf{A}) $ is a vector formed by the diagonal elements of $ \mathbf{A} $, and $ \mathrm{Diag}(\mathbf{a}) $ is a diagonal matrix with the entries of $ \mathbf{a} $ on its diagonal. 
$\mathbf{I}_N$ is the $ N\times N $ identity matrix; $ \lceil \cdot \rceil $ denotes the ceiling integer operation.
$ n \choose k $ denotes the $ k $-combinations taken from $ n $; $ a^+=\max\{0,a\} $;
$ \circ $ denotes the Khatri-Rao product; $ \otimes $ denotes the Kronecker product; and $ * $ denotes the Hadamard product.
For a set $\mathcal{A} $, $ |\mathcal{A}| $ denotes its cardinality. 
For two sets $\mathcal{A}$ and $ \mathcal{B} $, $\mathcal{A}\setminus\mathcal{B}$ denotes the difference of two sets, which consists of elements that are in $\mathcal{A}$ but not in $ \mathcal{B} $.
Expectation is denoted by $ \mathbb{E}[\cdot] $.

\section{System model}
\begin{figure}[htbp]
	\vspace{-0.5cm}
	\centering{\includegraphics[width=.5\textwidth]{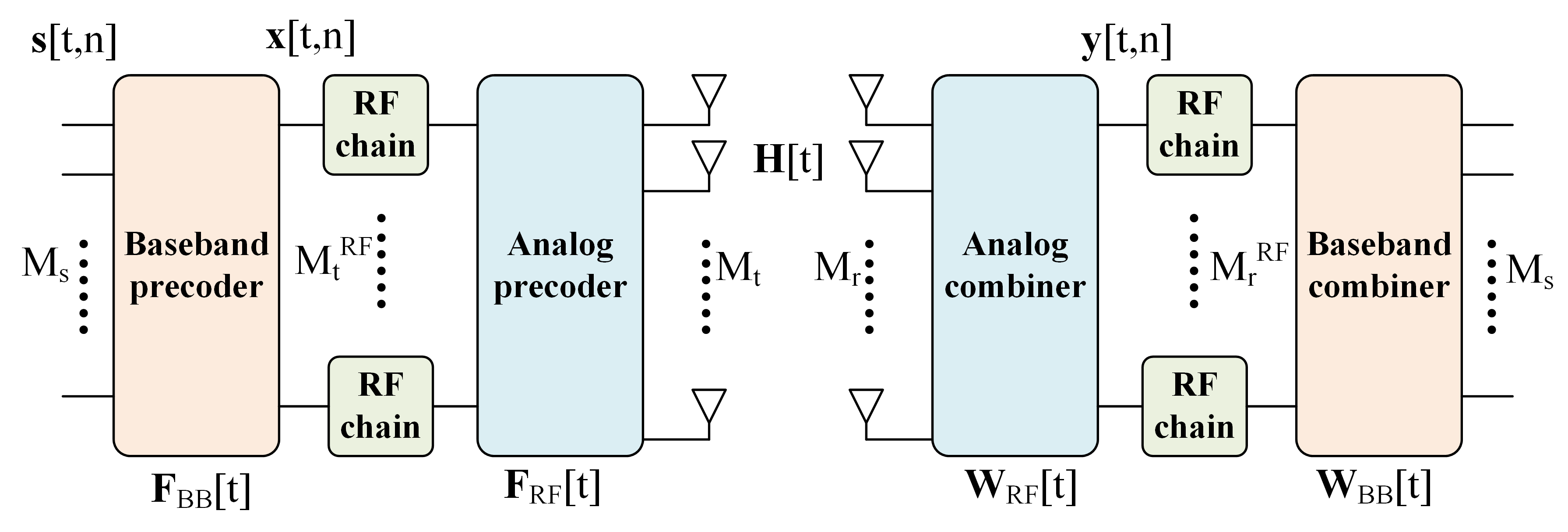}}  	
	\caption{MmWave massive MIMO communication with hybrid beamforming.} 
	\label{hybrid_architecture}  
\end{figure}

As shown in Fig. \ref{hybrid_architecture}, we consider a mmWave massive MIMO communication system, where a BS with $ M_t $ transmit antennas and $ M^\mathrm{RF}_t < M_t $ RF chains sends $ M_s $ data streams to a UE, which has $ M_r $ receive antennas and $ M^\mathrm{RF}_r < M_r $ RF chains. 
Without loss of  generality, we assume that $ M_r^\mathrm{RF}\leq M_t^\mathrm{RF} $, and $ M_s=\min\{M_r^\mathrm{RF}, M_t^\mathrm{RF}\}=M_r^\mathrm{RF} $, since the case of $ M_s<M_r^\mathrm{RF} $ can be treated as a special case by setting the power of the corresponding data streams to zero. 
We focus on the downlink communication, and the proposed technique can be similarly applied to the uplink communication.
The BS and UE apply baseband precoders/combiners $ \mathbf{F}_\mathrm{BB}\in \mathbb{C}^{M^\mathrm{RF}_t \times M_s} $ and $ \mathbf{W}_\mathrm{BB}\in \mathbb{C}^{M^\mathrm{RF}_r \times M_s} $, together with RF precoders/combiners $ \mathbf{F}_\mathrm{RF}\in \mathbb{C}^{M_t \times M^\mathrm{RF}_t} $ and $ \mathbf{W}_\mathrm{RF}\in \mathbb{C}^{M_r \times M^\mathrm{RF}_r} $, respectively.
Since the RF precoders/combiners $ \mathbf{F}_\mathrm{RF} $ and $ \mathbf{W}_\mathrm{RF} $ are implemented using analog phase shifters, their entries have constant modulus, i.e., $  |\mathbf{[F_\mathrm{RF}]}_{i,j}|=1/\sqrt{M_t},\ i=1,\ldots,M_t, \  j=1,\ldots,M^\mathrm{RF}_t $, and $  |\mathbf{[W_\mathrm{RF}]}_{i,j}|=1/\sqrt{M_r},\ i=1,\ldots,M_r, \  j=1,\ldots,M^\mathrm{RF}_r$.
Furthermore, we assume that codebook-based analog beamforming is applied, where the columns of $ \mathbf{F}_\mathrm{RF} $ and $ \mathbf{W}_\mathrm{RF} $ are selected from predetermined codebooks $ \mathcal{F} $ and $ \mathcal{W} $, respectively, and the total number of candidate beamforming vectors are denoted by $ \mathcal{|F}| $ and $ |\mathcal{W|} $, respectively. 
Thus, the total number of possible choices for $\mathbf{ F}_\mathrm{RF} $ is $ |\mathcal{F}|\choose M^\mathrm{RF}_t $, which are denoted as $ \mathbf{F}_{\mathrm{RF},i} $, where $  i=1,2,\ldots, {|\mathcal{F}|\choose M^\mathrm{RF}_t}$. 
Similarly, the total number of possible choices for $\mathbf{W}_\mathrm{RF} $ is $ |\mathcal{W}|\choose M^\mathrm{RF}_r $, which are denoted as $ \mathbf{W}_{\mathrm{RF},j} $, where $ j\in \{1,2,\ldots, {|\mathcal{W}|\choose M^\mathrm{RF}_r}\} $. 
On the other hand, once $ \mathbf{F}_\mathrm{RF} $ and $ \mathbf{W}_\mathrm{RF} $ are determined, the baseband precoders/combiners $ \mathbf{F}_\mathrm{BB} $ and $ \mathbf{W}_\mathrm{BB} $ are assumed to be freely designed since the resulting effective channel has much lower dimension.  
The transmit power constraint of the BS is met by satisfying $\left\|\mathbf{F}_{\mathrm{RF}} \mathbf{F}_{\mathrm{BB}}\right\|_{F}^{2}=1$.

We consider the quasi-static block-fading channel model, where the MIMO channel coefficients remain constant for each channel coherent block of $ N $ symbol durations, and may vary across different blocks.
For each coherent block $ t $, denote the MIMO channel matrix as $ \mathbf{H}[t]\in \mathbb C^{M_r\times M_t} $.
Further denote $ \mathbf{F}_\mathrm{RF}[t]$ and $  \mathbf{W}_\mathrm{RF}[t]$ as the selected analog beamforming matrices during coherent block $t$ at the BS and UE, respectively.
Therefore, over $ T $ channel coherent blocks, the resulting digital baseband input-output relationship can be written as
\begin{equation}
	\begin{aligned}
		\mathbf{y}[t,n]=\sqrt{P}\mathbf{W}_\mathrm{RF}^H[t]\mathbf{H}[t]\mathbf{F}_\mathrm{RF}[t]\mathbf{x}[t,n]+\mathbf{W}_\mathrm{RF}^H[t]\mathbf{n}[t,n], \
		n=1,2,\ldots,N,t=1,2,\ldots, T,
		\label{receive signal}
	\end{aligned}
\end{equation}
where $ \mathbf{y}[t,n]\in \mathbb{C}^{M_r^\mathrm{RF} \times 1}  $ is the $n$th symbol of coherent block $t$ after applying analog beamforming $ \mathbf{W}_\mathrm{RF}[t] $, but before applying digital baseband beamforming (as labelled in Fig. \ref{hybrid_architecture}), $P$ denotes the transmit power, $\mathbf{x}[t,n]\in \mathbb{C}^{M_t^\mathrm{RF}\times 1} $ is the $ n $th signal vector of coherent block $ t $ after baseband precoding, but before analog precoding, and $ \mathbf{n}[t,n]\in\mathbb{C}^{M_r\times 1} $ is the zero-mean circularly symmetric complex Gaussian (CSCG) noise vector with variance $  \sigma^2$, i.e., $ \mathbf{n}[t,n] \sim \mathcal{CN}(\mathbf{0},\sigma^2\mathbf{I}_{M_r}) $. 
Define the effective noise as $ \tilde{ \mathbf{n}}[t,n]=\mathbf{W}_\mathrm{RF}^H[t] \mathbf{n}[t,n]$, which is no longer white but follows the distribution $\tilde{\mathbf{n}}[t,n]\sim \mathcal{CN}(\mathbf 0, \sigma^2 \mathbf{W}_\mathrm{RF}^H[t] \mathbf{W}_\mathrm{RF}[t]) $. 
Thus, for any selected analog beamformers $ \mathbf{F}_\mathrm{RF}[t] $ and $\mathbf{ W}_\mathrm{RF}[t] $, (1) is essentially a fully digital MIMO communication system with colored noise, which can be converted to MIMO AWGN channel by noise whitening. Specifically, by multiplying (1) with $ (\mathbf{W}_\mathrm{RF}^{H}[t] \mathbf{W}_\mathrm{RF}[t])^{-\frac{1}{2}} $, we have
\begin{equation}
	\begin{aligned}
		\tilde{\mathbf{y}}[t,n]=\sqrt{P}\mathbf{H}_\mathrm{e}[t]\mathbf{x}[t,n]+\hat{\mathbf{n}}[t,n], \
		n=1,2,\ldots,N,t=1,2,\ldots, T,
		\label{noisewhiten}
	\end{aligned}
\end{equation}
where $ \mathbf{H}_\mathrm{e}[t]\in \mathbb{C}^{M_r^\mathrm{RF} \times M_t^\mathrm{RF}} $ is the equivalent baseband digital MIMO channel after noise whitening, given by 
\begin{equation}
	\begin{aligned}
	\mathbf{H}_\mathrm{e}[t] = (\mathbf{W}_\mathrm{RF}^{H}[t] \mathbf{W}_\mathrm{RF}[t])^{-\frac{1}{2}} \mathbf{W}_{\mathrm{RF}}^{H}[t] \mathbf{H}[t]\mathbf{F}_{\mathrm{RF}}[t],
	\label{He}
	\end{aligned}
\end{equation}
and $ \hat{\mathbf{n}}[t,n]\in\mathbb{C}^{M_r^\mathrm{RF}\times 1} $ is the zero-mean CSCG white noise vector with variance $  \sigma^2$, i.e., $\hat{ \mathbf{n}}[t,n] \sim \mathcal{CN}(\mathbf{0},\sigma^2\mathbf{I}_{M^\mathrm{RF}_r}) $.

For the equivalent baseband digital MIMO AWGN channel (\ref{noisewhiten}), it is well known that the capacity-achieving signalling scheme is the CSCG signaling \cite{goldsmith2005wireless}, i.e., $\mathbf{x}[t,n]\sim \mathcal{CN}(\mathbf{0}, \mathbf{R}_\mathrm{x}[t])$, where $ \mathbf{R}_\mathrm{x}[t]=\mathbb{E}[\mathbf{x}[t,n]\mathbf{x}^H[t,n]] $ denotes the covariance matrix for coherent block $t$. 
As a result, with sufficiently long block length $N$, the average  spectral efficiency in bits per second per Hertz (bps/Hz) over the $T$ coherent blocks is
\begin{equation}
	\begin{aligned}
	R=\frac{1}{T}\sum_{t=1}^{T} \log _{2} \left|\mathbf{I}_{M_r^\mathrm{RF}}+\tilde{P}\mathbf{H}_\mathrm{e}[t]\mathbf{R}_\mathrm{x}[t]\mathbf{H}_\mathrm{e}^H[t] \right|,
	\label{rate}
	\end{aligned}
\end{equation} 
where $ \tilde{P}=P/\sigma^2 $ denotes the transmit signal-to-noise ratio (SNR).

\section{Optimal Hybrid Beamforming with Perfect CSI}
To provide the performance upper bound for our considered system, in this section, under the assumption that the channel matrix $ \mathbf{H}[t] $ is perfectly known for each coherent block $t$, the optimal digital beamforming matrices $ \mathbf{F}_\mathrm{BB}[t] $ and $ \mathbf{W}_\mathrm{BB}[t] $ and the codebook-based analog beamforming matrices $\mathbf{F}_\mathrm{RF}[t]$ and $\mathbf{W}_\mathrm{RF}[t]$ are obtained to maximize the average spectral efficiency.
With the rate expression (\ref{rate}), it is noted that the RF precoders/combiners $ \mathbf{F}_\mathrm{RF}[t] $ and $ \mathbf{W}_\mathrm{RF}[t] $ mainly affect the effective digital MIMO channel $ \mathbf{H}_\mathrm{e}[t] $ via (\ref{He}), and the baseband precoder $\mathbf{F}_\mathrm{BB}[t]$ determines the covariance matrix $ \mathbf{R}_\mathrm{x}[t] $. 
Specifically, let $\mathbf{s}[t,n]\in \mathbb{C}^{M_s\times 1}$ be the information-bearing symbols of the $M_s$ data streams, which is i.i.d. CSCG distributed with zero mean and normalized power, i.e., $\mathbf{s}[t,n]\sim \mathcal{CN}(\mathbf 0, \mathbf{I}_{M_s})$. 
With $\mathbf{x}[t,n]=\mathbf{F}_\mathrm{BB}[t]\mathbf{s}[t,n]$, we have $\mathbf{R}_\mathrm{x}[t]=\mathbf{F}_\mathrm{BB}[t]\mathbf{F}_\mathrm{BB}^H[t]$, and the transmit power constraint $\|\mathbf{F}_\mathrm{RF}[t]\mathbf{F}_\mathrm{BB}[t]\|^2_F=1$ is thus equivalent to $\mathrm{tr}(\mathbf{F}_\mathrm{RF}^H[t]\mathbf{F}_\mathrm{RF}[t]\mathbf{R}_\mathrm{x}[t])=1$. 
As a result, to optimize the baseband precoder $\mathbf{F}_\mathrm{BB}[t]$, one only needs to optimize its covariance matrix $\mathbf{R}_\mathrm{x}[t]$.
Furthermore, it is not difficult to see that the hybrid beamforming optimization problem to maximize the average rate (\ref{rate}) can be decoupled for each channel coherent block, which is given by
\begin{equation}
	\begin{aligned}
	& \max_{\mathbf{F}_\mathrm{RF}\in \mathcal{F},\mathbf{W}_\mathrm{RF}\in \mathcal{W},\mathbf{R}_\mathrm{x}\succeq0} \ \log _{2} \left|\mathbf{I}_{M_r^\mathrm{RF}}+\tilde{P}\mathbf{H}_\mathrm{e}\mathbf{R}_\mathrm{x}\mathbf{H}_\mathrm{e}^H \right|,\\
		&\quad \quad \quad \mathrm{s.t.}\ \ \mathrm{tr}(\mathbf{F}_\mathrm{RF}^H\mathbf{F}_\mathrm{RF}\mathbf{R}_\mathrm{x})=1.\qquad\quad
		\label{problem}
	\end{aligned}
\end{equation} 
Note that in  the above formulation, we have suppressed the time index $t$ for notational simplicity. 
Furthermore, the baseband receive beamforming matrix $\mathbf{W}_\mathrm{BB}$ is not  directly involved in the optimization problem, since it can be readily obtained once the precoder and RF combiner are obtained, as will become clear later.
	
To find the optimal solution to (\ref{problem}), for any given analog beamforming matrices $ \mathbf{F}_\mathrm{RF}$ and $ \mathbf{W}_\mathrm{RF} $, we first find the optimal covariance matrix $ \mathbf{R}_\mathrm{x} $, and hence the corresponding baseband precoders $ \mathbf{F}_\mathrm{BB}$ and $ \mathbf{W}_\mathrm{BB}$. 
Based on such results, the globally optimal solution can then be obtained by exhaustively searching over all possible selections of $ \mathbf{F}_\mathrm{RF}\in \mathcal{F}$ and $ \mathbf{W}_\mathrm{RF}\in \mathcal{W}$.    

For any given analog beamforming matrices $ \mathbf{F}_\mathrm{RF}$ and $ \mathbf{W}_\mathrm{RF} $ and with the perfect CSI on channel matrix $ \mathbf{H} $, the effective channel matrix $ \mathbf{H}_\mathrm{e} $ is determined based on (\ref{He}). 
Therefore, the subproblem of (\ref{problem}) for baseband beamforming optimization is expressed as 
\begin{equation}
	\begin{aligned}
		\max_{\mathbf{R}_\mathrm{x}\succeq0}& \ \log _{2} \left|\mathbf{I}_{M_r^\mathrm{RF}}+\tilde{P}\mathbf{H}_\mathrm{e}\mathbf{R}_\mathrm{x}\mathbf{H}_\mathrm{e}^H \right|,\\
		\mathrm{s.t.}&\ \ \mathrm{tr}(\mathbf{F}_\mathrm{RF}^H\mathbf{F}_\mathrm{RF}\mathbf{R}_\mathrm{x})=1.\qquad\quad
		\label{optproblem}
	\end{aligned}
\end{equation} 
Problem (\ref{optproblem}) is similar to the standard MIMO transmit covariance matrix optimization problem \cite{goldsmith2005wireless}, but with a linear transformation $\mathbf{F}_\mathrm{RF}^H\mathbf{F}_\mathrm{RF}$ on $\mathbf{R}_\mathrm{x}$ for the power constraint.
By defining $\tilde{\mathbf{R}}_\mathrm{x}=(\mathbf{F}_\mathrm{RF}^H\mathbf{F}_\mathrm{RF})^{\frac{1}{2}}\mathbf{R}_\mathrm{x}(\mathbf{F}_\mathrm{RF}^H\mathbf{F}_\mathrm{RF})^{\frac{1}{2}} $, so that $ \mathbf{R}_\mathrm{x} =(\mathbf{F}_\mathrm{RF}^H\mathbf{F}_\mathrm{RF})^{-\frac{1}{2}}\tilde{\mathbf{R}}_\mathrm{x}(\mathbf{F}_\mathrm{RF}^H\mathbf{F}_\mathrm{RF})^{-\frac{1}{2}}$, problem (\ref{optproblem}) can be converted to 
\begin{equation}
	\begin{aligned}
		\max_{\tilde{\mathbf{R}}_x\succeq0}&\ \log _{2} \left|\mathbf{I}_{M_r^\mathrm{RF}}+\tilde{P}\tilde{\mathbf{H}}\tilde{\mathbf{R}}_\mathrm{x}\tilde{\mathbf{H}}^H \right|\\
		\mathrm{s.t.}&\ \ \mathrm{tr}(\tilde{\mathbf{R}}_x)=1,
		\label{optimizationproblem}
	\end{aligned}
\end{equation} 
where
\begin{equation}
	\begin{aligned}
		\tilde{\mathbf{H}}=\mathbf{H}_\mathrm{e}(\mathbf{F}_\mathrm{RF}^H\mathbf{F}_\mathrm{RF})^{-\frac{1}{2}}
		=(\mathbf{W}_\mathrm{RF}^H\mathbf{W}_\mathrm{RF})^{-\frac{1}{2}}\mathbf{W}_\mathrm{RF}\mathbf{H}\mathbf{F}_\mathrm{RF}(\mathbf{F}_\mathrm{RF}^H\mathbf{F}_\mathrm{RF})^{-\frac{1}{2}},
		\label{Htilde}
	\end{aligned}
\end{equation}
where the last equality follows from (\ref{He}). Problem (\ref{optimizationproblem}) is the standard MIMO capacity optimization problem, whose optimal solution is known to be the eigenmode transmission \cite{goldsmith2005wireless}. 
Specifically, by applying singular value decomposition (SVD) to the equivalent channel matrix $ \tilde{\mathbf{H}}\in \mathbb{C}^{M_r^\mathrm{RF}\times M_t^\mathrm{RF}}$, we have
\begin{equation}
	\begin{aligned}
		\tilde{\mathbf{H}}=&\mathbf{U} \mathbf{\Sigma} \mathbf{V}^{H}
		=&\mathbf{U}\left[\mathbf{\Sigma}_1\quad\mathbf{0}\right]\left[\begin{array}{l}
			\mathbf{V}_{1}^{H}\\
			\mathbf{V}_{2}^{H}
		\end{array}\right]
		=&\mathbf{U} \mathbf{\Sigma}_1 \mathbf{V}_1^{H},
		\label{SVD}
	\end{aligned}
\end{equation}
where $ \mathbf{U}\in \mathbb{C}^{M^\mathrm{RF}_r \times M^\mathrm{RF}_r}  $ and $ \mathbf{V}\in \mathbb{C}^{M^\mathrm{RF}_t \times M^\mathrm{RF}_t}  $ are unitary matrices, $ \mathbf{\Sigma}=[\mathbf{\Sigma}_1, \mathbf{0}] \in \mathbb{R}^{M^\mathrm{RF}_r \times M^\mathrm{RF}_t} $, with $ \mathbf{\Sigma}_1\in \mathbb{R}^{M^\mathrm{RF}_r \times M^\mathrm{RF}_r}$ being a diagonal matrix with the diagonal elements giving by the non-zero singular values $ \sigma_{1} \geq \sigma_{2} \geq \dots \geq \sigma_{M_r^\mathrm{RF}}>0$.
$\mathbf{V}_1\in \mathbb{C}^{ M_t^\mathrm{RF}\times M_r^\mathrm{RF}} $ is the submatrix of $ \mathbf{V} $.
By applying  the Hadamard's inequality for positive semi-definite matrix, the optimal solution to (\ref{optimizationproblem}) is
\begin{equation}
	\tilde{\mathbf{R}}^\mathrm{opt}_x=\mathbf{V}_1\mathbf{\Gamma}\mathbf{V}_1^H,
	\label{Rxopt}
\end{equation}
where $\mathbf{ \Gamma}=\mathrm{Diag}\{\rho_1,\ldots,\rho_{M_r^\mathrm{RF}}\} $ is the diagonal matrix of power allocation coefficient, which is given by the classic water-filling solution.
As a result, the optimal transmit covariance matrix is
\begin{equation}
	\mathbf{R}_\mathrm{x}=(\mathbf{F}_\mathrm{RF}^{H} \mathbf{F}_\mathrm{RF})^{-\frac{1}{2}}\mathbf{V}_1\mathbf{\Gamma}\mathbf{V}_1^H(\mathbf{F}_\mathrm{RF}^{H} \mathbf{F}_\mathrm{RF})^{-\frac{H}{2}},
\end{equation}
and the corresponding baseband precoder/combiner matrix $ \mathbf{F}_\mathrm{BB} $/$ \mathbf{W}_\mathrm{BB}$ can be determined as 
\begin{equation}
	\begin{aligned}
		&\mathbf{F}_\mathrm{BB}=\mathbf{R}_\mathrm{x}^\frac{1}{2}=(\mathbf{F}_\mathrm{RF}^{H} \mathbf{F}_\mathrm{RF})^{-\frac{1}{2}}\mathbf{V}_1\mathbf{ \Gamma}^{\frac{1}{2}},\
		\mathbf{W}_\mathrm{BB}= (\mathbf{W}_\mathrm{RF}^{H} \mathbf{W}_\mathrm{RF})^{-\frac{1}{2}}\mathbf{U}.
		\label{Basebandbeamforming}
	\end{aligned}
\end{equation}
Note that the factor $ (\mathbf{W}_\mathrm{RF}^H\mathbf{W}_\mathrm{RF})^{-
\frac{1}{2}} $ for $ \mathbf{W}_\mathrm{BB} $ accounts for the noise whitening matrix introduced in (\ref{noisewhiten}), and $ \mathbf{U} $ is used to diagonalize the equivalent MIMO AWGN channel based on (\ref{SVD}). 

It follows from (\ref{Basebandbeamforming}) that for any given analog beamforming matrices $ \mathbf{F}_\mathrm{RF} $/$ \mathbf{W}_\mathrm{RF}$, the optimal baseband beamformig matrices can be obtained in closed-form. 
Therefore, the optimal solution to (\ref{optproblem}) can be found by exhaustively searching over all the possible selections of $ \mathbf{F}_\mathrm{RF} $ and $ \mathbf{W}_\mathrm{RF} $ from the codebooks $ \mathcal{F}$ and $\mathcal{W} $, which is summarized in Algorithm 1.
\begin{algorithm}[t]
	\caption{Optimal Hybrid Beamforming with Codebook-Based Analog Beamforming} 
	\hspace*{0.02in} {\bf Input:} 
	The channel matrix $\mathbf{H}$, analog beamforming codebooks $\mathcal{F}$ and $\mathcal{W}$\\
	\hspace*{0.02in} {\bf Output:} 
	The optimal hybrid beamforming matrices $\mathbf{F}_\mathrm{RF}^\mathrm{opt},\mathbf{W}_\mathrm{RF}^\mathrm{opt},\mathbf{F}_\mathrm{BB}^\mathrm{opt},\mathbf{W}_\mathrm{BB}^\mathrm{opt}$
	\begin{algorithmic}[1]
		\State Initialize $ R_\mathrm{max}=0 $
		\For{$ i=1,\ldots,{|\mathcal{F}|\choose M^\mathrm{RF}_t}$} 
		\State $ \mathbf{F}_\mathrm{RF} = \mathbf{F}_{\mathrm{RF},i} $
		\For{$ j=1,\ldots,{|\mathcal{W}|\choose M^\mathrm{RF}_r}$} 
		\State $ \mathbf{W}_\mathrm{RF} = \mathbf{W}_{\mathrm{RF},j} $
		\State Obtain communication $ R $
		\If{$ R>R_\mathrm{max} $}
		\State $ \mathbf{F}_\mathrm{RF}^\mathrm{opt}=\mathbf{F}_\mathrm{RF}, \mathbf{W}_\mathrm{RF}^\mathrm{opt}=\mathbf{W}_\mathrm{RF} $
		\State Obtain $ \mathbf{F}_\mathrm{BB} $ and $ \mathbf{W}_\mathrm{BB} $ based on (\ref{Basebandbeamforming}), and let
		\Statex \qquad \quad \quad \ $ \mathbf{F}_\mathrm{BB}^\mathrm{opt}=\mathbf{F}_\mathrm{BB}, \mathbf{W}_\mathrm{BB}^\mathrm{opt}=\mathbf{W}_\mathrm{BB}, R_\mathrm{max}=R $
		\EndIf
		\EndFor
		\EndFor
		\State \Return $\mathbf{F}_\mathrm{RF}^\mathrm{opt},\mathbf{W}_\mathrm{RF}^\mathrm{opt},\mathbf{F}_\mathrm{BB}^\mathrm{opt},\mathbf{W}_\mathrm{BB}^\mathrm{opt}$
	\end{algorithmic}
\end{algorithm}

As shown in Algorithm 1, to obtain the optimal beamforming solutions, an exhaustive search of $ { |\mathcal{F}|}\choose {M_{t}^{\mathrm{RF}}}$ ${ |\mathcal{W}|}\choose{M_{r}^{\mathrm{RF}}}$ combinations is needed, and each requires SVD over the equivalent digital MIMO channel of dimension $ M_r^\mathrm{RF}\times M_t^\mathrm{RF} $. 
As the SVD typically requires complexity $ \mathcal{O}\left(\left({M_{r}^{\mathrm{RF}}}\right)^3\right) $ \cite{watkins2004fundamentals}, the overall computational complexity is $ \mathcal{O}\left({{ |\mathcal{F}|}\choose{M_{t}^{\mathrm{RF}}}}{{ |\mathcal{W}|}\choose{M_{r}^{\mathrm{RF}}}}\left({M_{r}^{\mathrm{RF}}}\right)^3\right) $.
Therefore, while Algorithm 1 gives the global optimal solution, it incurs prohibitive high computation cost for mmWave massive MIMO system even when $ M_t^\mathrm{RF} $ and/or $ M_r^\mathrm{RF} $ are moderately large.
Thus, Algorithm 1 can only be used to obtain the performance upper bound for small-size MIMO systems. 

Note that Algorithm 1 also requires the availability of perfect CSI. In practice, CSI needs to be acquired via e.g., pilot-based training, which incurs high training overhead for mmWave massive MIMO systems with 
limited RF chains.
Specifically, for each coherent block $ t $, let $ N_\mathrm{tr}<N $ denote the number of symbol durations used for channel training.
In order to estimate the channel matrix $ \mathbf{H}[t] $ with dimension $ M_r \times M_t $ for each coherent block $ t $, at least $ M_r M_t $ measurements are required for meaningful estimation.
With $  M_r^\mathrm{RF} $ RF chains at  the UE, at least $ N_\mathrm{tr}\geq \lceil\frac{M_rM_t}{M_r^\mathrm{RF}}\rceil$ training durations are required for each coherent block, which is prohibitive when $ M_rM_t\gg M_r^\mathrm{RF} $. 
While some techniques like CS have been proposed to reduce the training overhead, they rely on additional assumptions on e.g. channel spatial sparsity and perfect knowledge of array geometry and orientations, and also require high computational complexity due to the sophisticated iterative algorithms for signal processing.
Note that with the channel training-based hybrid beamforming design, it follows from (\ref{rate}) that the effective average communication rate can be written as
\begin{align}
		R=&\frac{1}{T}\sum_{t=1}^{T}\frac{N-N_{tr}}{N} \log _{2} \left|\mathbf{I}_{M_r^\mathrm{RF}}+\tilde{P}\hat{\mathbf{H}}_\mathrm{e}[t]\hat{\mathbf{R}}_x[t]\hat{\mathbf{H}}_\mathrm{e}^H[t] \right|
		\leq&\left(1-\frac{N_{t r}}{N}\right)R_\mathrm{opt}
		\leq&\left(1-\frac{M_{r} M_{t}}{NM_r^\mathrm{RF}}\right) R_\mathrm{opt},
		\label{trainrate}
\end{align}
where $ \hat{\mathbf{H}}_\mathrm{e}[t] $ denotes the effective digital MIMO channel based on (\ref{He}), for which the analog beamforming matrices $\hat{ \mathbf{F}}_\mathrm{RF}[t] $/$ \hat{\mathbf{W}}_\mathrm{RF}[t] $ and covariance matrices $ \hat{ \mathbf{R}}_\mathrm{x}[t] $ are obtained based on the imperfectly estimated channel $ \hat{\mathbf{H}}[t] $, and $ R_\mathrm{opt}  $ denotes maximum rate from Algorithm 1 under perfect CSI assumption, and the pre-log factor $ \frac{N-N_{t r}}{N} $ accounts for the training overhead.
The first inequality of (\ref{trainrate}) is due to the fact that the covariance matrix $  \hat{\mathbf{R}}_\mathrm{x}[t] $ and the corresponding analog/digital beamforming matrices are determined based on the estimated channel matrix $ \hat{\mathbf{H}}[t] $ rather than the true channel $ \mathbf{H}[t] $;
the second inequality follows since $ N_\mathrm{tr}\geq \lceil\frac{M_rM_t}{M_r^\mathrm{RF}}\rceil\geq  \frac{M_rM_t}{M_r^\mathrm{RF}}$ is typically required.
The result in (\ref{trainrate}) shows that for large MIMO systems with $ M_rM_t $ comparable to or even exceeding $ NM_r^\mathrm{RF} $ (e.g., $ N=1500,\ M_r=16,\ M_t=256,\ M_r^\mathrm{RF}=2 $, so that the pre-log factor is 0.32), there is a huge performance gap between the rate achieved by the channel training-based beamforming and the maximum rate $ R_\mathrm{opt} $ with perfect CSI.
Besides the prohibitive training overhead, another critical challenge for channel-training based beamforming lies in the prohibitive computation required for calculating the optimal beamforming matrices, as shown in Algorithm 1.

To address the above issues of high training overhead and high computational complexity, in this paper, we propose an environment-aware based hybrid beamforming scheme enabled by CKM.
The proposed technique utilizes two important information that critically impacts the actual communication channels: namely the UE location, which is more readily available in contemporary wireless systems with improving accuracy, and CKM, which is a site-specific database that contains useful location-specific channel-related information for all potential UE locations of interest and reveal the actual radio propagation environment.

\section{Environment-Aware Hybrid Beamforming via channel knowledge map}
\subsection{Key Idea of CKM-Enabled Environment-Aware Communications}
Before presenting the details of our proposed CKM-enabled environment-aware hybrid beamforming technique, it is useful to take a fresh new look at wireless communication channels.
At an abstract level, a wireless channel is mainly determined by the radio wave property (such as wavelength), the transmitter/receiver locations, and the actual radio propagation environment.
Over the past few decades, numerous efforts have been devoted to the mathematical characterization of wireless channels using stochastic and/or geometric-based approaches \cite{doc}. 
However, such channel modeling approaches only utilize the partial information of the transmitter/receiver locations (such as the link distance only, rather than their exact locations) and the very coarse environment information (such as urban, suburban, or rural area only, rather than the exact environment where the communication takes place). 
While such modeling approaches are tractable and easy for generalization, their modeled channels inevitably incur non-negligible errors when applied in actual communication scenarios, which thus necessitates the real-time channel estimation via pilot-based channel training. 
On the other hand, with the continuous advancement of localization technologies and enhanced environment awareness of UEs, environment-aware wireless communication is promising to resolve the issue of prohibitive training overhead in large-dimension MIMO systems \cite{CKM}. 

\begin{figure}[htbp] 				
	\vspace{-0.5cm}
	\centering{\includegraphics[width=.5\textwidth]{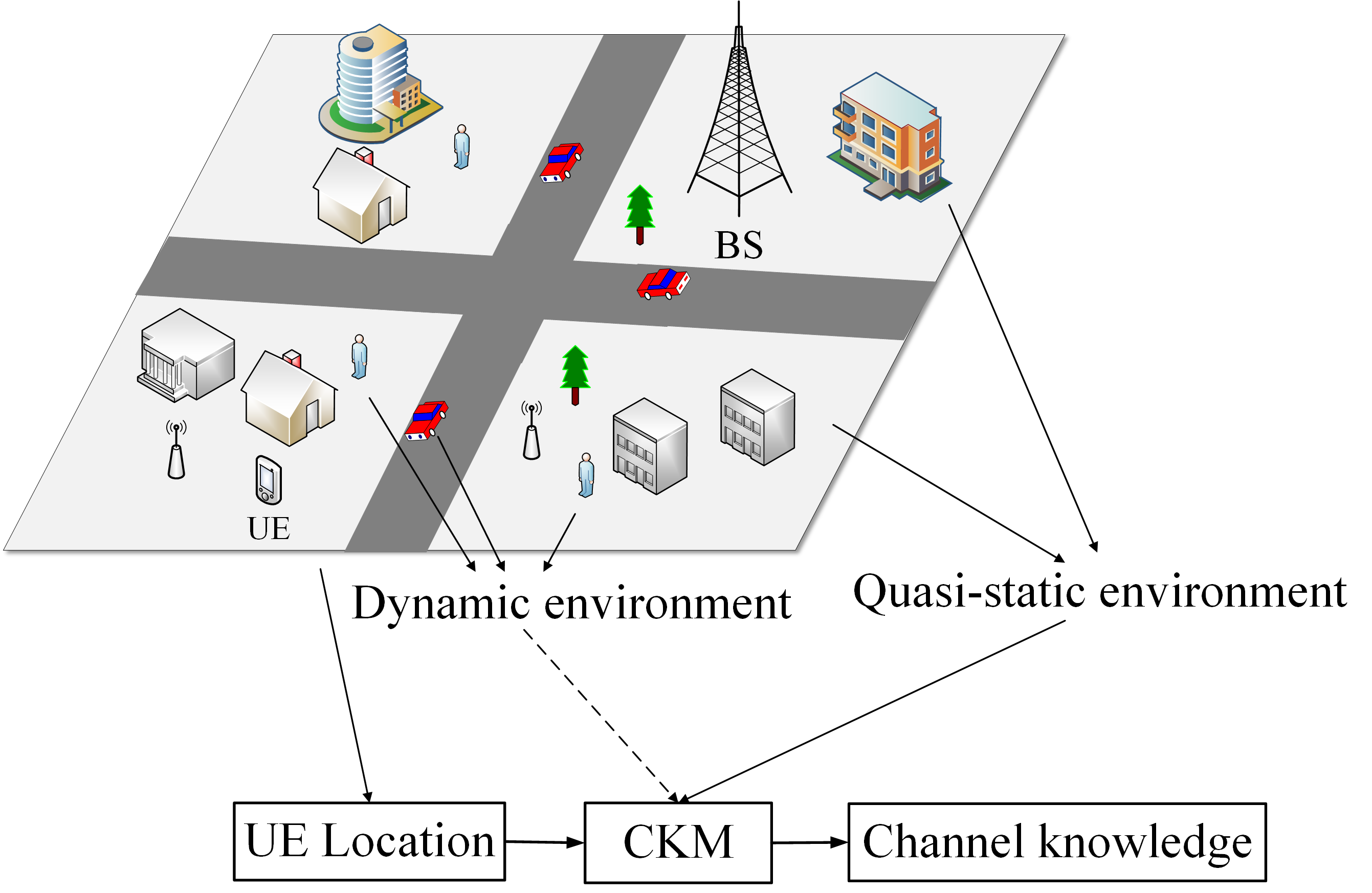}}  	 		 		
	\caption{An illustration of CKM for enabling environment-aware communications. } 
	\label{CKMgraph}  	 		 	
\end{figure}

For our considered mmWave massive MIMO communication system, with the BS location being fixed, the variation of channel matrix $  \mathbf{H}[t] $ is mainly due to the change of the UE location, which is denoted as $\mathbf {q}[t]$, as well as the variation of the actual communication environment, which is denoted as $E[t]$. Therefore, a generic representation of the MIMO channel matrix $ \mathbf{H}[t] $ can be written as 
\begin{equation}
	\mathbf{H}[t]=g_1(\mathbf{q}[t], E[t]),  \ t=1,2,....,T,
	\label{channelmodel}
\end{equation}
where $g_1(.,.)$ is an arbitrary function. In fact, though various approximations are possible, such as those based on stochastic models, it is very difficult, if not impossible, to obtain accurate expressions for $g_1(.,.)$ in complex environment, mainly due to the difficulty in mathematically modeling the environment $ E[t] $ accurately, as well as the sophisticated interaction between radio waves and actual environment. Fortunately, such an issue can be circumvented by leveraging the useful concept of CKM \cite{CKM}.
As illustrated in Fig. \ref{CKMgraph}, the influence of the environment on the channel is captured by the CKM, whereby the site-specific channel knowledge can be obtained with the knowledge of the UE location.

One straightforward type of CKM is the channel matrix map (CMM), which directly provides the channel coefficient matrix $\hat {\mathbf H}[t]$ for any given UE location $\mathbf q[t]$. 
Specifically, let $\mathcal Q$ denote the location space that includes all potential UE locations of interest, i.e., $\mathbf q[t]\in \mathcal Q$, $\forall t$. 
Then a CMM, which is denoted as $\mathcal {M}_\mathrm{CMM}[t]$, is a mapping $\mathbf q[t]\in \mathcal Q \rightarrow \hat {\mathbf H}[t]$. 
The map $\mathcal{M}_\mathrm{CMM}[t]$ is time-dependent in general, accounting for the time variation of the surrounding radio propagation environment, while excluding the impact of the location variation of the UE itself since that is captured by its trajectory over time $ \{\mathbf{q}[t]\} $.
Fortunately, in most practical scenarios, the wireless propagation environment (such as the locations, heights, and dielectric properties of the surrounding objects such as walls and buildings) does not change or changes in a much larger time scale than the UE location variations, as illustrated in Fig. \ref{CKMgraph}. 
Furthermore, even for those environment factors (such as pedestrians acting as scatterers) that may vary with comparable time scale as UE locations, the impact of the former on wireless channels is usually much less than the latter in practice.
Thus, the update of the CKM, in general, is necessary only when there is significant environment change, which is over a much larger time scale than the channel coherent time due to the UE location variation. 
As a result, provided that sufficiently accurate UE locations $\{\mathbf q[t]\}_{t=1}^T$ are available (say via GPS or cellular-based localization technologies), the channel matrix $  \hat {\mathbf{H}}[t] $ can be in principle obtained with the CMM $\mathcal {M}_\mathrm{CMM}[t]$, without or with very limited real-time channel training required.

However, a CMM that aims to directly retrieve the MIMO channel coefficients based on UE locations requires excessive storage resources. 
To address this issue, the concept of channel path map (CPM) was proposed in the preliminary version of this work \cite{wu2021environment}.
CPM aims to offer location-specific channel path information (such as the number of significant paths and their power, phases, and AoAs/AoDs) for all potential UE locations. 
Compared with CMM, CPM could be utilized to reconstruct the MIMO channel matrix based on the essential path information with much reduced storage requirement.
This is motivated by the classical geometry-based channel model, but with the fine location-specific  (instead of the coarse site-specific) channel path parameters, which is given by  
\begin{equation}
	\begin{aligned}			 
		\mathbf{H}[t]=\sqrt{M_r M_t} \sum^{L[t]}_{l=1} \alpha_{l}[t]
		\mathbf{a}_r(\theta_l[t]) \mathbf{a}_t^H(\phi_l[t]),\ t=1,...,T,
		\label{channel}
	\end{aligned}	 	
\end{equation}
where $ L[t] $ is the number of significant channel paths at channel coherent block $ t $, $ \alpha_{l}[t],  \theta_l[t]$, and $ \phi_l[t]$ are the complex gain, AoA, and AoD of the $l$th path, respectively. 
In general, the AoA and AoD contain both zenith and azimuth directions, i.e., $ \theta_l[t]=(\theta^z_l[t],\theta^a_l[t]) $, and $ \phi_l[t]=(\phi^z_l[t],\phi^a_l[t]) $, with the superscript $z$ and $a$ denoting the zenith and azimuth directions, respectively. 
Furthermore, $ \mathbf{a}_t(.) $ and $ \mathbf{a}_r(.) $ denote the transmit and receive array response vectors, respectively.
As a result, provided that the array configurations and the channel path information is known for each potential UE location, the channel matrix $ \mathbf{H}[t] $ can be reconstructed based on (\ref{channel}).
To this end, CPM plays the role of mapping the UE location to the essential path information \cite{wu2021environment}.
However, one limitation of  CPM lies in its difficulty to adapt to the dynamic environment. Specifically, once a CPM is constructed so that the channel matrix $ \mathbf{H}[t] $ can be inferred from the UE location $\mathbf{q}[t]$, the impact of environment variation can only be accounted for by updating the complete CPM, which is costly and inconvenient in practice .

To overcome the above limitations, in the following, we propose two new types of CKM, namely CAM and BIM, which render environment-aware communication possible even for dynamic environment with only light real-time training required.         

\subsection{CAM-Enabled Hybrid Beamforming}
\begin{figure}[htbp]
	\vspace{-0.5cm}
	\centering{\includegraphics[width=.50\textwidth]{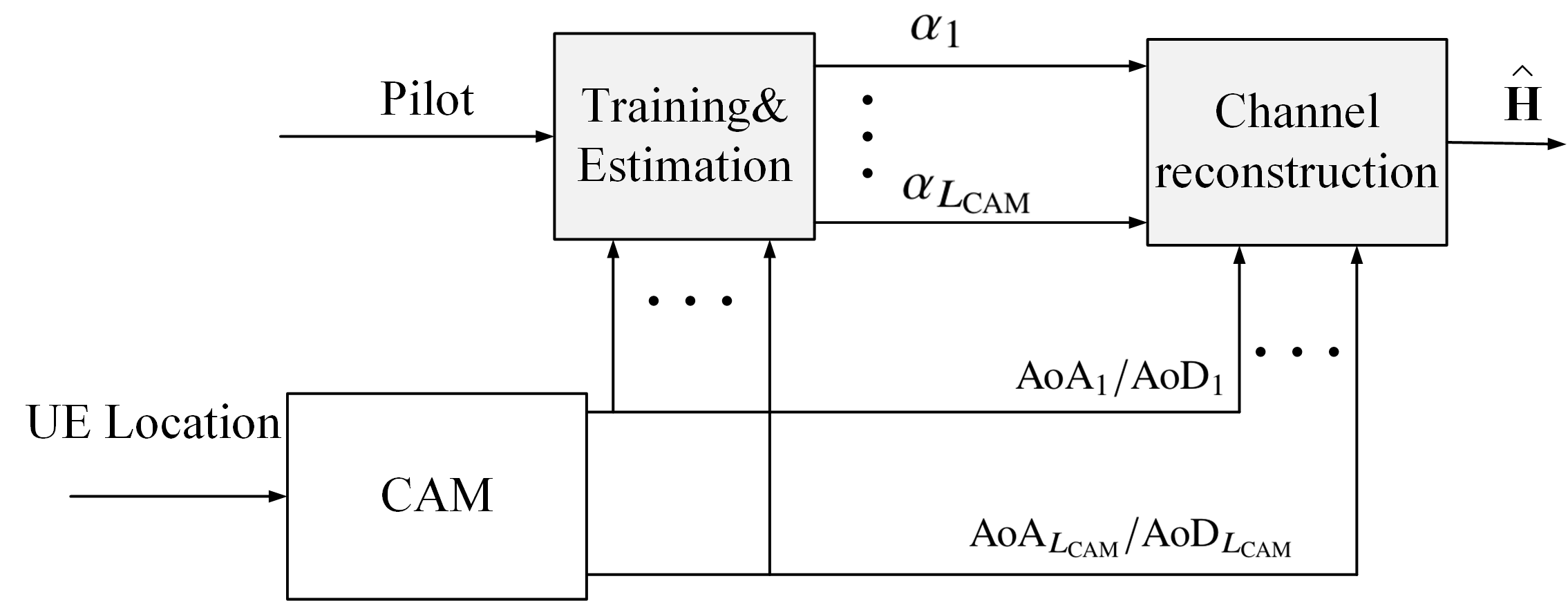}} 
	\caption{An illustration of CAM-enabled environment-aware hybrid beamforming with light training.} 
	\label{CAM3}  
\end{figure}

Different from CPM  that aims to predict all essential information of channel paths at each potential UE location, CAM mainly concerns about the path angle information, which is regarded as a large-scale parameter \cite{akdeniz2014millimeter}, while leaving the flexibility to further train the small-scale fading coefficient such as the path gains, to cater to environment variations. 
This is illustrated in Fig. \ref{CAM3} and elaborated in the following.

To illustrate the concept of CAM, we first express the channel model in (\ref{channel}) in a compact form as
\begin{equation}
	\mathbf{H}[t]=\mathbf{A}_r(\Theta[t])\mathrm{Diag}(\bm{\alpha}[t])\mathbf{A}_t^H(\Phi[t]),
	\label{channelcompact}
\end{equation}
where $ \bm{\alpha}[t]=\sqrt{M_rM_t}\left[\alpha_1[t],\alpha_2[t],\ldots,\alpha_{L[t]}[t]\right] ^T$, and
$ \Theta[t] $ and $ \Phi[t] $ are ordered tuples of AoAs and AoDs, respectively, i.e.,
\begin{equation}
	\Theta[t]=\{\theta_l[t]|l=1,\ldots,L[t]\}, \Phi[t]=\{\phi_l[t]|l=1,\ldots,L[t]\}.
\end{equation}
Besides, $ \mathbf{A}_r(\Theta[t]) \in \mathbb{C}^{M_r\times L}$ and $ \mathbf{A}_t(\Phi[t]) \in \mathbb{C}^{M_t\times L}$ are matrices given by
\begin{equation}
	\mathbf{A}_r(\Theta[t])=\left[\mathbf{a}_r(\theta_1[t])\ \ldots\ \mathbf{a}_r(\theta_{L[t]}[t])\right], 	\mathbf{A}_t(\Phi[t])=\left[\mathbf{a}_t(\phi_1[t])\ \ldots\ \mathbf{a}_r(\phi_{L[t]}[t])\right].
	\label{Ar}
\end{equation}
Furthermore, we define the set of groundtruth ordered tuple of AoAs and AoDs as 
\begin{equation}
	\begin{aligned}			 
		\Omega[t]=\{( \theta_l[t],\phi_l[t])|l=1,\ldots,L[t]\}.
	\end{aligned}	 	
\end{equation}

On the other hand, depending on the spatial resolution provided by the transmit and receive antenna arrays, the complete AoAs and AoDs in the space can be discretized into grids of discretized angles \cite{alkhateeb2014channel,lee2016channel}. 
Specifically, let $ J_r $ denote the number of discretized angles for azimuth AoA, so that $\theta_j^a \in \{0, 2\pi/J_r,\ldots, 2\pi(J_r-1)/J_r\}$, $ j=1,\ldots,J_r $. 
Similarly, let  $ I_r, I_t$ and $J_t$ denote the number of discretized angles for $ \theta_i^z $, $ \phi_i^z $ and $ \phi_j^a $.
Thus, the complete tuple of AoAs and AoDs can be written as 
\begin{equation}
	\bar{\Theta}=\{(\theta^z_i,\theta^a_j)|i=1,\ldots,I_r,j=1,\ldots,J_r\}, 	\bar{\Phi}=\{(\phi^z_i,\phi^a_j)|i=1,\ldots,I_t,j=1,\ldots,J_t\}.
\end{equation}
Therefore, the complete set of order pairs of all possible paths can be denoted as $ 	\bar{\Omega}=\bar{\Theta}\times\bar{\Phi} $, which has the cardinality of $ |\bar{\Omega}|=I_rJ_rI_tJ_t$.
Therefore, for each channel coherent block with spatial channel sparsity, the groundtruth channel angle pair $ \Omega[t]$ is only a subset of the complete angle tuple $ \bar{\Omega} $ , i.e., $ \Omega[t] \subset \bar \Omega $.

Furthermore, determining the groundtruth angle subset $ \Omega[t]$ from $ \bar{\Omega} $ can be regarded as a sparse recovery problem, which can be solved with techniques such as CS off-line.
To reduce the angle quantization error, the number of grids should be larger than the antenna size \cite{estimate}, i.e. $ I_rJ_r\geq M_r,  I_tJ_t\geq M_t $, which still results in significant training overhead and computational complexity even if CS is used.
Taking the orthogonal matching pursuit (OMP) in \cite{lee2016channel} as an example, it needs $\mathcal{O}(\frac{L[t]\ln( |\bar{\Omega}|)}{M_r^\mathrm{RF}}) $ measurements and $ \mathcal{O}(L[t] |\bar{\Omega}|\ln( |\bar{\Omega}|)) $ computations, which is still a prohibitively large number in massive MIMO systems.
Besides, without any prior information on the channel angle tuple $ \Omega[t]$, the observation matrix design for CS is aimless and may need large number of measurements to achieve a notable observation gain.


To address the above issues, we propose the concept of CAM to achieve environment-aware beamforming.
CAM aims to provide the location-specific candidate AoA/AoD tuples for all possible UE locations of interest. 
Specifically, CAM is a mapping from the UE location $ \mathbf{q}[t]\in \mathcal{Q} $ to the candidate AoA and AoD set as 
\begin{equation}
	\mathbf{q}[t]\in \mathcal{Q} \rightarrow  \hat{\Omega}[t],\quad t=1,2,....,T.
	\label{CAMmap}
\end{equation}
where $ \hat{\Omega}[t]\subset\bar{\Omega} $ denotes the candidate angle tuple with $ \hat{L}[t] $ elements for UE located at $ \mathbf{q}[t] $, which includes the AoAs/AoDs of $ \hat{L}[t] $ potential paths, i.e., 
\begin{equation}
	\begin{aligned}			 
		\hat{\Omega}[t]=\{( \theta_l[t],\phi_l[t])|l=1,\ldots,\hat{L}[t]\}, \ t=1,...,T.
		\label{CAM2}
	\end{aligned}	 	
\end{equation}

Note that compared to the complete angle tuple $\bar{\Omega}$, the location-specific candidate angle tuple $\hat{\Omega}[t]$ predicted by the CAM is expected to have much smaller size, which can thus be used to significantly reduce the real-time training overhead to reconstruct the channel in (\ref{channelcompact}).  
In the ideal case when $\hat \Omega[t]= \Omega[t]$, i.e., the CAM gives perfect AoA/AoD information so that the array response matrices $ \mathbf{A}_r(..)  $ and $ \mathbf{A}_t(..) $ in (\ref{channelcompact}) are perfectly known, then it only requires to perform real-time training to estimate the channel path gains $\bm{\alpha}[t]$ in order to reconstruct the channel $ \mathbf{H}[t] $ in (\ref{channelcompact}). 
For the practical scenario when $\hat \Omega[t] \neq \Omega[t]$, we have the following two cases. 
In the following, the time index $t$ is omitted for notational simplicity.

1) $ \Omega \subset \hat{\Omega} $:  In this case, the groundtruth channel angle tuple $ \Omega $ is a subset of that predicted by CAM. As a result, the groundtruth channel in (\ref{channelcompact}) can be expressed based on $ \hat{\Omega} $  as 
\begin{equation}
	\begin{aligned}
		\mathbf{H}=&\mathbf{A}_{r}(\hat{\Theta})\mathrm{Diag}(\hat{\bm{\alpha}})\mathbf{A}_{t}^H(\hat{\Phi}),
		\label{channelCAM1}
	\end{aligned}
\end{equation}
where $ \hat{\bm{\alpha}} $ is the extended vector including the groundtruth path gain $  \bm{\alpha} $ as well as zeros for those paths with path angles $ \hat{\Omega}\setminus\Omega$. 
Based on the prior knowledge of $ \mathbf{A}_r(\hat{\Theta}) $ and $\mathbf{A}_t(\hat{\Phi})$, we only need to estimate the path gains $ \hat{\bm{\alpha}} $, which can be achieved by light training, as elaborated in the following. 

For each channel coherence block $ t $, we first assume that a training interval of $ M_s $ symbol durations is used to train the path gain $ \hat{\bm{\alpha}} $. 
Let $ \mathbf S \in \mathbb{C}^{M_s \times M_s} $ denote the pilot symbols sent during the $ M_s $ symbol durations. 
Then, the resulting received signal matrix $ \mathbf{Y} \in \mathbb{C}^{M_s\times M_s}$ by concatenating the signals over $ M_s $ symbol durations can be written as
\begin{equation}
	\begin{aligned}
		\mathbf{Y}=\sqrt{P}\mathbf{W}_\mathrm{BB}^H\mathbf{W}_\mathrm{RF}^H\mathbf{H}\mathbf{F}_\mathrm{RF}\mathbf{F}_\mathrm{BB}\mathbf{S}+\mathbf{W}_\mathrm{BB}^H\mathbf{W}_\mathrm{RF}^H\mathbf{N},
		\label{Ytraining}
	\end{aligned}
\end{equation}
where $ \mathbf{W}_\mathrm{BB}$,  $\mathbf{W}_\mathrm{RF}$, $ \mathbf{F}_\mathrm{RF}$, and $\mathbf{F}_\mathrm{BB} $ are the beamforming matrices during the training phase, which are determined later, $ \mathbf{N} \in \mathbb{C}^{M_s\times M_s}$ is the noise matrix. Let $ \tilde{\mathbf{N}} = \mathbf{W}_\mathrm{BB}^H\mathbf{W}_\mathrm{RF}^H\mathbf{N}$ for notational simplicity.
We assume that orthonormal pilot sequences are used, so that $\mathbf{S}\mathbf{S}^H=\mathbf{I}_{M_s} $. After projecting $ \mathbf{Y} $ into $\mathbf{S}^H$ and taking the vectorization, we have
\begin{align}
		\tilde{\mathbf{y}}=&\mathrm{vec}(\mathbf{Y}\mathbf{S}^H)\notag\\
		=&\mathrm{vec}(\sqrt{P}\mathbf{W}_\mathrm{BB}^H\mathbf{W}_\mathrm{RF}^H\mathbf{H}\mathbf{F}_\mathrm{RF}\mathbf{F}_\mathrm{BB})+\mathrm{vec}(\tilde{\mathbf{N}}\mathbf{S}^H)\notag\\
		=&\sqrt{P}\mathrm{vec}(\mathbf{W}_\mathrm{BB}^H\mathbf{W}_\mathrm{RF}^H\mathbf{A}_{r}(\hat{\Theta})\mathrm{Diag}(\hat{\bm{\alpha}})\mathbf{A}_{t}^H(\hat{\Phi})\mathbf{F}_\mathrm{RF}\mathbf{F}_\mathrm{BB})+\mathrm{vec}(\tilde{\mathbf{N}}\mathbf{S}^H)\notag\\
		=&\sqrt{P}\left((\mathbf{F}_\mathrm{BB}^T\mathbf{F}_\mathrm{RF}^T\mathbf{A}_{t}^{*}(\hat{\Phi}))\circ( \mathbf{W}_\mathrm{BB}^H\mathbf{W}_\mathrm{RF}^H\mathbf{A}_{r}(\hat{\Theta}))\right)
		\hat{\bm{\alpha}}+\mathrm{vec}(\tilde{\mathbf{N}}\mathbf{S}^H)\notag\\
		=&\sqrt{P}\hat{\mathbf{Q}}\hat{\bm{\alpha}}+\mathrm{vec}(\tilde{\mathbf{N}}\mathbf{S}^H),
		\label{CAMY}
\end{align}
where $ \hat{\mathbf{Q}}\in \mathbb{C}^{M^2_s\times\hat{L}} $ is the observation matrix, given by 
\begin{equation}
	\begin{aligned}
 \hat{\mathbf{Q}}=( \mathbf{F}_\mathrm{BB}^T\mathbf{F}_\mathrm{RF}^T\mathbf{A}_{t}^{*}(\hat{\Phi}))\circ( \mathbf{W}_\mathrm{BB}^H\mathbf{W}_\mathrm{RF}^H\mathbf{A}_{r}(\hat{\Theta})).
	\end{aligned}
\label{Qhat}
\end{equation}
Note that the third equality of (\ref{CAMY}) follows from (\ref{channelCAM1}), and the forth equality follows from the properties of the Khatri-Rao product vec($\mathbf{A}\text{Diag}(\mathbf{a})\mathbf{B}$)=$(\mathbf{B}^T\circ\mathbf{A})\mathbf{a}$ \cite{trees2002optimum}. 

With (\ref{CAMY}), if $ M_s^2\geq\hat L $ so that the observation matrix $  \hat{\mathbf{Q}} $ has full column rank, the path gain $ \hat{\bm{\alpha}} $ can be estimated as
\begin{equation}
	\begin{aligned}
		\hat{\hat{\bm{\alpha}}}=\frac{1}{\sqrt{P}}(\hat{\mathbf{Q}}^H\hat{\mathbf{Q}})^{-1}\hat{\mathbf{Q}}^H\tilde{\mathbf{y}}
	=\hat{\bm{\alpha}}+\frac{1}{\sqrt{P}}(\hat{\mathbf{Q}}^H\hat{\mathbf{Q}})^{-1}\hat{\mathbf{Q}}^H\mathrm{vec}(\tilde{\mathbf{N}}\mathbf{S}^H),
		\label{LS}
	\end{aligned}
\end{equation}
and the mean square error (MSE) is given by 
\begin{equation}
	\begin{aligned}
		\mathrm{MSE}=&\mathbb{E}\left[\left| \left| \frac{1}{\sqrt{P}}(\hat{\mathbf{Q}}^H\hat{\mathbf{Q}})^{-1}\hat{\mathbf{Q}}^H\mathrm{vec}(\tilde{\mathbf{N}}\mathbf{S}^H)\right| \right| _2^2\right]
		=\mathbb{E}\left[\left| \left| \frac{1}{\sqrt{P}}(\hat{\mathbf{Q}}^H\hat{\mathbf{Q}})^{-1}\hat{\mathbf{Q}}^H\mathrm{vec}(\tilde{\mathbf{N}})\right| \right| _2^2\right].
		\label{MSE}
	\end{aligned}
\end{equation}
where the second equality is due to that $ \mathbf{S} $ is a unitary matrix and will not affect the distribution of  $ \tilde{\mathbf{N}} $.
From (\ref{MSE}), a well designed observation matrix $ \hat{\mathbf{Q}} $ can reduce the MSE, so here we give a possible observation matrix design under the assumption that $ \mathbf{W}_\mathrm{BB}^H\mathbf{W}_\mathrm{RF}^H \mathbf{W}_\mathrm{RF}\mathbf{W}_\mathrm{BB}=\mathbf{I}_{M_s}$, i.e., $ \mathrm{vec}(\tilde{\mathbf{N}})\sim \mathcal{CN}(\mathbf{0},\sigma^2\mathbf{I}_{M_s^2}) $.

{\it Lemma 1:} If $ \mathrm{vec}(\tilde{\mathbf{N}})\sim \mathcal{CN}(\mathbf{0},\sigma\mathbf{I}_{M_s^2}) $, the MSE in (\ref{MSE}) can be rewritten as 
\begin{equation}
	\mathrm{MSE}\geq \frac{\sigma^2\hat{L}^2}{P||\mathbf{A}_t^H(\hat{\Phi})\mathbf{F}_\mathrm{RF}\mathbf{F}_\mathrm{BB}||_2^F||\mathbf{A}_r^H(\hat{\Theta})\mathbf{W}_\mathrm{RF}\mathbf{W}_\mathrm{BB}||_2^F}.
\end{equation}

{\it Proof:} Please refer to Appendix.

Then the observation matrix design problem can be converted to 
\begin{equation}
	\begin{aligned}
		(\mathbf{F}_\mathrm{BB},\mathbf{F}_\mathrm{RF})&=\mathop{\arg\max}_{\mathbf{F}_\mathrm{BB},\mathbf{F}_\mathrm{RF}} \left| \left| \mathbf{A}_t^H(\hat{\Phi})\mathbf{F}_\mathrm{RF}\mathbf{F}_\mathrm{BB}\right| \right| ^2_F,\\
		\mathrm{s.t.}&\ \ \mathrm{tr}(\mathbf{F}_\mathrm{BB}^H\mathbf{F}_\mathrm{RF}^H\mathbf{F}_\mathrm{RF}\mathbf{F}_\mathrm{BB})=1,
	\end{aligned}
	\label{OF}
\end{equation}
and 
\begin{equation}
	\begin{aligned}
		(\mathbf{W}_\mathrm{BB},\mathbf{W}_\mathrm{RF})&=\mathop{\arg\max}_{\mathbf{W}_\mathrm{BB},\mathbf{W}_\mathrm{RF}} \left| \left| \mathbf{A}_r^H(\hat{\Theta})\mathbf{W}_\mathrm{RF}\mathbf{W}_\mathrm{BB}\right| \right| ^2_F,\\
		\mathrm{s.t.}&\ \
		\mathbf{W}_\mathrm{BB}^H\mathbf{W}_\mathrm{RF}^H \mathbf{W}_\mathrm{RF}\mathbf{W}_\mathrm{BB}=\mathbf{I}_{M_s}.
	\end{aligned}
	\label{OW}
\end{equation}
The optimization problem (\ref{OF}) and (\ref{OW}) can be solved by first designing the analog beamforming matrices. 
Let $ p_{1},\ldots, p_{M_t^\mathrm{RF}}\in \{1,...,|\mathcal{F}|\}  $ denote the $ M_t^\mathrm{RF} $ selected beam indices from $ \mathcal{F} $, so that the analog transmit beamforming matrices is $ \mathbf{F}_\mathrm{RF}=[\mathbf{f}_{p_1},\ldots,\mathbf{f}_{p_{M_t^\mathrm{RF}}}] $. 
Similarly, let $ q_1,\ldots, q_{M_r^\mathrm{RF}}\in \{1,...,|\mathcal{W}|\}  $ denote the $ M_r^\mathrm{RF} $ selected beam indices from $ \mathcal{W} $, so that the analog receive beamforming matrices is $ \mathbf{W}_\mathrm{RF}=[\mathbf{w}_{q_1},\ldots,\mathbf{w}_{q_{M_r^\mathrm{RF}}}] $.
Then, the analog beamforming matrices can be determined by maximizing the Frobenius norm in analog domain as 
\begin{equation}
	\begin{aligned}
		\mathbf{F}_\mathrm{RF}=&\mathop{\arg\max}_{\mathbf{F}_\mathrm{RF}\in\mathcal{F}} \left| \left| \mathbf{A}_t^H(\hat{\Phi})\mathbf{F}_\mathrm{RF}\right| \right| ^2_F
		=\mathop{\arg\max}_{p_1,\ldots, p_{M_t^\mathrm{RF}} \in \{1,2,\ldots,|\mathcal{F}|\}}\sum_{m=1}^{M_t^\mathrm{RF}} \left| \left| \mathbf{A}_t^H(\hat{\Phi})\mathbf{f}_{p_m}\right| \right| ^2_2,\\
			\mathbf{W}_\mathrm{RF}=&\mathop{\arg\max}_{\mathbf{W}_\mathrm{RF}\in\mathcal{F}} \left| \left| \mathbf{A}_r^H(\hat{\Theta})\mathbf{W}_\mathrm{RF}\right| \right| ^2_F
			=\mathop{\arg\max}_{q_1,\ldots, q_{M_r^\mathrm{RF}} \in \{1,2,\ldots,|\mathcal{W}|\}}\sum_{n=1}^{M_r^\mathrm{RF}} \left| \left| \mathbf{A}_r^H(\hat{\Theta})\mathbf{w}_{q_n}\right| \right| ^2_2,
	\end{aligned}
\end{equation}
which can be solved via a search over $ \mathcal{F} $ and $ \mathcal{W} $.
By applying eigenvalue decomposition (EVD) on $ \mathbf{F}_\mathrm{RF}^H \mathbf{A}_t(\hat{\Phi})\mathbf{A}_t^H(\hat{\Phi}) \mathbf{F}_\mathrm{RF}$ and $\mathbf{W}_\mathrm{RF}^H \mathbf{A}_r(\hat{\Theta})\mathbf{A}_r^H(\hat{\Theta}) \mathbf{W}_\mathrm{RF}$, we have 
\begin{equation}
	\begin{aligned}
		  &\mathbf{F}_\mathrm{RF}^H \mathbf{A}_t(\hat{\Phi})\mathbf{A}_t^H(\hat{\Phi}) \mathbf{F}_\mathrm{RF}=\hat{\mathbf{V}} \mathbf{\Sigma}_t \hat{\mathbf{V}}^{H}
		=\left[\hat{\mathbf{V}}_{1}\quad\hat{\mathbf{V}}_{2}\right]\left[\begin{array}{l}
			\mathbf{\Sigma}_{t_1} \ \mathbf{0}\\
			\mathbf{0}  \quad\mathbf{\Sigma}_{t_1}
		\end{array}\right]\left[\begin{array}{l}
			\hat{\mathbf{V}}_{1}^{H} \\
			\hat{\mathbf{V}}_{2}^{H}
		\end{array}\right],\\
		&\mathbf{W}_\mathrm{RF}^H \mathbf{A}_r(\hat{\Theta})\mathbf{A}_r^H(\hat{\Theta}) \mathbf{W}_\mathrm{RF}=\hat{\mathbf{U}} \mathbf{\Sigma}_r \hat{\mathbf{U}}^{H},
		\label{SVDFW}
	\end{aligned}
\end{equation}
where $\hat{ \mathbf{V}}\in \mathbb{C}^{M^\mathrm{RF}_t \times M^\mathrm{RF}_t}  $ and $ \hat{\mathbf{U}}\in \mathbb{C}^{M^\mathrm{RF}_r \times M^\mathrm{RF}_r}  $ are unitary matrices, $\hat{\mathbf{V}}_1\in \mathbb{C}^{ M_t^\mathrm{RF}\times M_s} $ is the submatrix of $ \hat{\mathbf{V}} $, $ \mathbf{\Sigma}_ t\in \mathbb{R}^{M^\mathrm{RF}_t \times M^\mathrm{RF}_t} $ and $ \mathbf{\Sigma}_r \in \mathbb{R}^{M^\mathrm{RF}_r \times M^\mathrm{RF}_r} $ are diagonal matrices containing the eigenvalues, $ \mathbf{\Sigma}_{t_1}\in \mathbb{R}^{M_s \times M_s}$ is a diagonal matrix with the diagonal elements giving by the $ M_s $ largest non-zero eigenvalues in $ \mathbf{\Sigma}_ t$.
Then, to allocate sufficient power to each path, equal power transmission is implemented, and the corresponding baseband precoder/combiner can be determined as
\begin{equation}
	\begin{aligned}
		\mathbf{F}_\mathrm{BB}=&\frac{1}{\sqrt{M_s}}(\mathbf{F}_\mathrm{RF}^{H} \mathbf{F}_\mathrm{RF})^{-\frac{1}{2}}\hat{\mathbf{V}}_1,\
		\mathbf{W}_\mathrm{BB}=& (\mathbf{W}_\mathrm{RF}^{H} \mathbf{W}_\mathrm{RF})^{-\frac{1}{2}}\hat{\mathbf{U}}.
	\end{aligned}
\end{equation}

On the other hand, for scenarios when $ \hat L> M_s^2 $, a total of  $ \lceil\frac{\hat{L}}{M_s}\rceil $ training epochs, each with $ M_s $ symbol durations can be used, and similar result as above can be obtained. 
After $ \hat{\hat{\bm{\alpha}}} $ being obtained in (\ref{LS}), together with the predicted angle information $ 
\hat{\Omega} $ from CAM, the MIMO channel matrix $ \mathbf{H} $ can be reconstructed based on (\ref{channelCAM1}), and beamforming matrices can be obtained via e.g. Algorithm 1 in Section III.

It is worth remarking that the proposed CAM-enabled environment-aware technique can also be applied together with other existing techniques, such as the CS-based channel estimation.
Specifically, in the conventional CS method, without any prior information, the observation matrix is obtained based on the complete angle grid $ \bar{\Omega} $, which can be written as $ \bar{\mathbf{Q}}=( \mathbf{F}_\mathrm{BB}^T\mathbf{F}_\mathrm{RF}^T\mathbf{A}_{t}^{*}(\bar{\Phi}))\otimes( \mathbf{W}_\mathrm{BB}^H\mathbf{W}_\mathrm{RF}^H\mathbf{A}_{r}(\bar{\Theta}))$, and training beamforming matrices are designed to minimize total coherence of $ \bar{\mathbf{Q}} $ (a sum of squared inner products of different columns in $ \bar{\mathbf{Q}} $) \cite{sung2020compressed}.
With CAM, the location-specific angle information $ \hat \Omega $ can be obtained, so that the more effective observation matrix $ \hat{\mathbf{Q}} $ in (\ref{Qhat}) can be designed to direct training power better match the channel angles.
Besides, with $ \hat L<<|\Omega| $, the measurements and computational complexity of the OMP method in \cite{lee2016channel} can be reduced to $\mathcal{O}(\frac{L\ln(\hat{L})}{M_r^\mathrm{RF}}) $ and $ \mathcal{O}(L\hat{L}\ln(\hat{L})) $, due to the reduction of the searching space for channel reconstruction.

2) $ \Omega\not \subset \hat \Omega $:
Next, we consider the case when the angles $ \hat \Omega $ predicted by CAM do not include all the groundtruth angles $ \Omega $.
Let $ \tilde{\Omega}= \Omega\setminus\hat{\Omega}$, $ \tilde{\Theta} $ and $ \tilde{\Phi} $ are the corresponding tuples of AoA and AoD in $ \tilde \Omega $. 
Then the channel matrix in (\ref{channelcompact}) can be decomposed as 
\begin{equation}
	\begin{aligned}
			\mathbf{H}=&\mathbf{A}_{r}(\hat{\Theta})\mathrm{Diag}(\hat{\bm{\alpha}})\mathbf{A}_{t}^H(\hat{\Phi})+\mathbf{A}_{r}(\tilde{\Theta})\mathrm{Diag}(\tilde{\bm{\alpha}})\mathbf{A}_{t}^H(\tilde{\Phi}),
		\label{channelCAM2}
	\end{aligned}
\end{equation}
where $ \tilde{\bm{\alpha}} $ is the corresponding path gain of those paths in $ \tilde{\Omega} $.
With the similar training procedures as (\ref{Ytraining}), the resulting signal in (\ref{CAMY}) can be expressed as 
\begin{equation}
	\begin{aligned}
		\tilde{\mathbf{y}}=\sqrt{P}(\hat{\mathbf{Q}}\hat{\bm{\alpha}}+\tilde{\mathbf{Q}}\tilde{\bm{\alpha}})+\mathrm{vec}(\tilde{\mathbf{N}}\mathbf{S}^H),
		\label{CAMY2}
	\end{aligned}
\end{equation}
where $ \tilde{\mathbf{Q}}=( \mathbf{F}_\mathrm{BB}^T\mathbf{F}_\mathrm{RF}^T\mathbf{A}_{t}^{*}(\tilde{\Phi}))\circ( \mathbf{W}_\mathrm{BB}^H\mathbf{W}_\mathrm{RF}^H\mathbf{A}_{r}(\tilde{\Theta})) $. 
Note that since only $ \hat{\mathbf{Q}} $ is known, while $ \tilde{\mathbf{Q}} $ is not, the path gain estimation is given by 
\begin{equation}
	\begin{aligned}
		\tilde{\hat{\bm{\alpha}}}=\frac{1}{\sqrt{P}}(\hat{\mathbf{Q}}^H\hat{\mathbf{Q}})^{-1}\hat{\mathbf{Q}}^H\tilde{\mathbf{y}}
		=\hat{\bm{\alpha}}+\frac{1}{\sqrt{P}}(\hat{\mathbf{Q}}^H\hat{\mathbf{Q}})^{-1}\hat{\mathbf{Q}}^H\tilde{\mathbf{Q}}\tilde{\bm{\alpha}}+\frac{1}{\sqrt{P}}(\hat{\mathbf{Q}}^H\hat{\mathbf{Q}})^{-1}\hat{\mathbf{Q}}^H\mathrm{vec}(\tilde{\mathbf{N}}\mathbf{S}^H).
		\label{LS2}
	\end{aligned}
\end{equation}

It can be seen from (\ref{LS2}) that the estimated path gain $ \tilde{\hat{\bm{\alpha}}} $ consists of two parts: one is the gains of those paths included in CAM $ \hat{\Omega} $, the other is the projection of the path gain of the unpredicted paths $ \tilde{\Omega} $ on that of $ \hat{\Omega} $.
Thus, even if the predicted angles $ \hat \Omega $ by CAM do not include all the groundtruth paths in $ \Omega $, the impact of these paths can also be partially reconstructed, and prediction accuracy can be improved by increasing the number of predicted paths by CAM, i.e., $ \hat{L} $.
Based on the estimated path gain in (\ref{LS2}) and the angle information $ 
\hat{\Omega} $ from CAM, the MIMO channel can be reconstructed based on (\ref{channelCAM1}).

Note that compared with the prevalent environment-unaware channel estimation methods, the CAM-enabled channel estimation only requires a real-time pilot training to estimate the gains of those candidate paths, based on the UE location information.
Besides, it may also lead to higher SNR in the training phase, since the precoding/combining matrices can be more effectively designed based on the location-specific candidate angle subset. 
Furthermore, the effective communication rate becomes
\begin{equation}
	\begin{aligned}
		R=&\frac{1}{T}\sum_{t=1}^{T}\frac{N-N_{tr}}{N} \log _{2} \left|\mathbf{I}_{M_r^\mathrm{RF}}+\tilde{P}\hat{\mathbf{H}}_\mathrm{e}[t]\hat{\mathbf{R}}_x[t]\hat{\mathbf{H}}_\mathrm{e}^H[t] \right|\\
		\leq&\left(1-\frac{N_{t r}}{N}\right)R_\mathrm{opt}\\
		\leq&\left(1-\frac{1}{N}\lceil\frac{\hat{L}[t]}{M_r^\mathrm{RF}}\rceil\right) R_\mathrm{opt},
		\label{CAMrate}
	\end{aligned}
\end{equation}
where the first inequality is due to the discrepancy between the reconstructed channel and the true channel in general;
the second inequality follows since $ N_\mathrm{tr}\geq \lceil\frac{\hat{L}[t]}{M_r^\mathrm{RF}}\rceil$.
It is observed from  (\ref{trainrate}) and (\ref{CAMrate}) that the large performance gap of the environment-unaware training-based hybrid beamforming can be significantly reduced by the CAM-enabled environment-aware design, since the number of unknown parameters is greatly reduced from $ M_rM_t $ to $ \hat{L}[t] $.

\subsection{BIM Enabled Hybrid Beamforming}
\begin{figure}[htbp]
	\vspace{-0.5cm}
	\centering{\includegraphics[width=.45\textwidth]{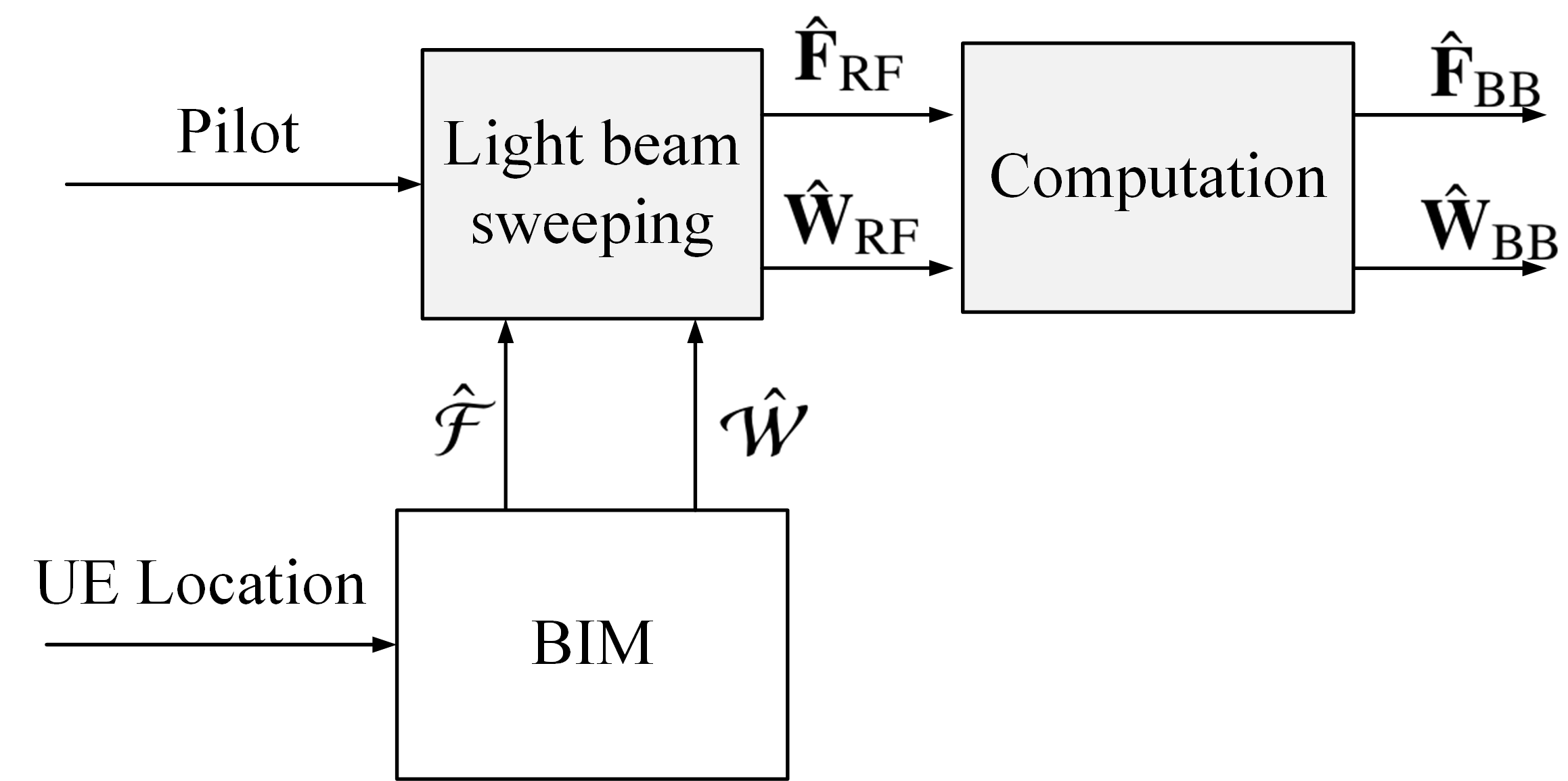}} 
	\caption{An illustration of BIM-enabled environment-aware hybrid beamforming.}
	\label{BIM3}  
\end{figure}

Note that the CAM-based approach still needs to reconstruct the MIMO channel before finding the best beamforming matrices.
Based on the discussions in Section III, though CAM is able to significantly save the real-time training overhead, finding the best beamforming matrices with the obtained MIMO channel matrices still incurs high computational complexity.  
To address this issue, in this subsection, we propose an alternative CKM-enabled hybrid beamforming scheme, based on BIM.
Different from CAM, BIM aims to provide the location-specific candidate transmit and receive analog beamforming vectors directly for all possible UE locations of interest, which can be specified by the following mapping:
\begin{equation} 
	\mathbf{q}[t]\in \mathcal{Q} \rightarrow  	(\hat{\mathcal{F}}[t],\hat{\mathcal{W}}[t]),\ \ t=1,2,....,T,
\end{equation}
where $ \hat{\mathcal{F}}[t]=\{\mathbf{f}_p|p=1,\ldots,|\hat{\mathcal{F}}[t]|\}\subset \mathcal{F}$ and $ \hat{\mathcal{W}}[t]=\{\mathbf{w}_q|q=1,\ldots,|\hat{\mathcal{W}}[t]|\}\subset \mathcal{W}$ are the subsets of the analog codebooks with significantly reduced cardinality, i.e., $M^\mathrm{RF}_t \leq |\hat{\mathcal{F}}[t]| \ll |\mathcal{F}| $ and $M^\mathrm{RF}_r \leq |\hat{\mathcal{W}}[t]|  \ll  |\mathcal{W}| $.
For notational simplicity, the time index $t$ is omitted in the following.
With the location-specific candidate analog beamforming vectors in $ \hat{\mathcal{F}} $ and $\hat {\mathcal{W}}$, the analog beamforming matrices $ \hat{\mathbf{F}}_\mathrm{RF}$ and $ \hat{\mathbf{W}}_\mathrm{RF}$ can be determined with only light beam sweeping, based on which the digital beamforming matrices $ \hat{\mathbf{F}}_\mathrm{BB}$ and $ \hat{\mathbf{W}}_\mathrm{BB}$ can be further obtained.
The key idea of BIM-enabled hybrid beamforming is illustrated in Fig. \ref{BIM3}.

Let $ p_1,\ldots, p_{M_t^\mathrm{RF}}\in \{1,...,|\hat{\mathcal{F}}|\}  $ denote the $ M_t^\mathrm{RF} $ selected beam indices from $ \hat{\mathcal{F}} $, so that the analog transmit beamforming matrix is denoted as $ \mathbf{F}_\mathrm{RF}=[\mathbf{f}_{p_1},\ldots,\mathbf{f}_{p_{M_t^\mathrm{RF}}}] $. 
Similarly, let $ q_1,\ldots, q_{M_r^\mathrm{RF}}\in \{1,...,|\hat{\mathcal{W}}|\}  $ denote the $ M_r^\mathrm{RF} $ selected beam indices from $ \hat{\mathcal{W}} $, so that the analog receive beamforming matrix is $ \mathbf{W}_\mathrm{RF}=[\mathbf{w}_{q_1},\ldots,\mathbf{w}_{q_{M_r^\mathrm{RF}}}] $.
One approach for determining $\mathbf{ F}_\mathrm{RF} $ and $\mathbf{W}_\mathrm{RF}$ is to select the analog beam vectors so that the resulting effective MIMO channel matrix has the largest sum-power gain, i.e.,
%
\begin{equation}
	\begin{aligned}
			(\mathbf{F}_\mathrm{RF},\mathbf{W}_\mathrm{RF})=\arg \max ||\mathbf{W}_\mathrm{RF}^H \mathbf{H} \mathbf{F}_\mathrm{RF}||_F^2 
			=\mathop{\arg\max}_{\substack{p_1,\ldots, p_{M_t^\mathrm{RF}} \in \{1,2,\ldots,|\hat{\mathcal{F}}|\}\\q_1,\ldots, q_{M_r^\mathrm{RF}} \in \{1,2,\ldots,|\hat{\mathcal{W}}|\}}} \sum_{m=1}^{M_t^\mathrm{RF}}\sum_{n=1}^{M_r^\mathrm{RF}} |\mathbf{w}_{q_n}^H\mathbf{H}\mathbf{f}_{p_m}|^2.			
	\end{aligned}
	\label{maxFW}
\end{equation}
Note that directly solving (\ref{maxFW}) not only requires knowing the channel matrix $ \mathbf{H} $, but also suffers from prohibitive computational complexity. 
To resolve such issues, we propose a light beam sweeping method, so that the resulting channel power for all possible beam pairs $ (\mathbf{f}_{p_m},\mathbf{w}_{q_n}) $ can be directly measured, based on which the beam selection in (\ref{maxFW}) can be obtained.
To this end, during the training phase, to evaluate the impact of analog beam selection only, the digital beamforming matrices are set as
\begin{equation}
	\begin{aligned}
		\mathbf{F}_\mathrm{BB}=\rho\left[\begin{array}{l}
			\mathbf{I}_{M_r^\mathrm{RF}} \\
			\mathbf{0}_{M_t^\mathrm{RF}-M_r^\mathrm{RF}}
		\end{array}\right],\
		\mathbf{W}_\mathrm{BB}=\mathbf{I}_{M_r^\mathrm{RF}},
	\end{aligned}
\label{FBBWBB}
\end{equation}
where $ \rho $ is the normalization factor to ensure that the transmit power constraint is satisfied with equality.
Besides, the beamforming vectors in $ \hat{\mathcal{F}} $ and $ \hat{\mathcal{W}} $ are partitioned into $ \lceil\frac{|\hat{\mathcal{F}}|}{M_r^\mathrm{RF}}\rceil$ and $ \lceil\frac{|\hat{\mathcal{W}}|}{M_r^\mathrm{RF}}\rceil $ groups, each with $ M_r^\mathrm{RF} $ vectors (except the last group). 
Therefore, the $i$th transmit and $j$th receive analog beamforming matrices are respectively denoted as
\begin{equation}
	\begin{aligned}
		\hat{\mathbf{F}}_\mathrm{RF,i}=&\left[\mathbf{f}_{(i-1)M_r^{\mathrm{RF}}+1},\ldots,\mathbf{f}_{iM_r^{\mathrm{RF}}}\right], i=1,\ldots,\lceil\frac{|\hat{\mathcal{F}}|}{M_r^\mathrm{RF}}\rceil,\\
		\hat{\mathbf{W}}_\mathrm{RF,j}=&\left[\mathbf{w}_{(j-1)M_r^{\mathrm{RF}}+1},\ldots,\mathbf{w}_{jM_r^{\mathrm{RF}}}\right], j=1,\ldots,\lceil\frac{|\hat{\mathcal{W}}|}{M_r^\mathrm{RF}}\rceil.
	\end{aligned}
\label{hatFW}
\end{equation}

\begin{figure}[htbp]
	\vspace{-0.5cm}
	\centering{\includegraphics[width=.50\textwidth]{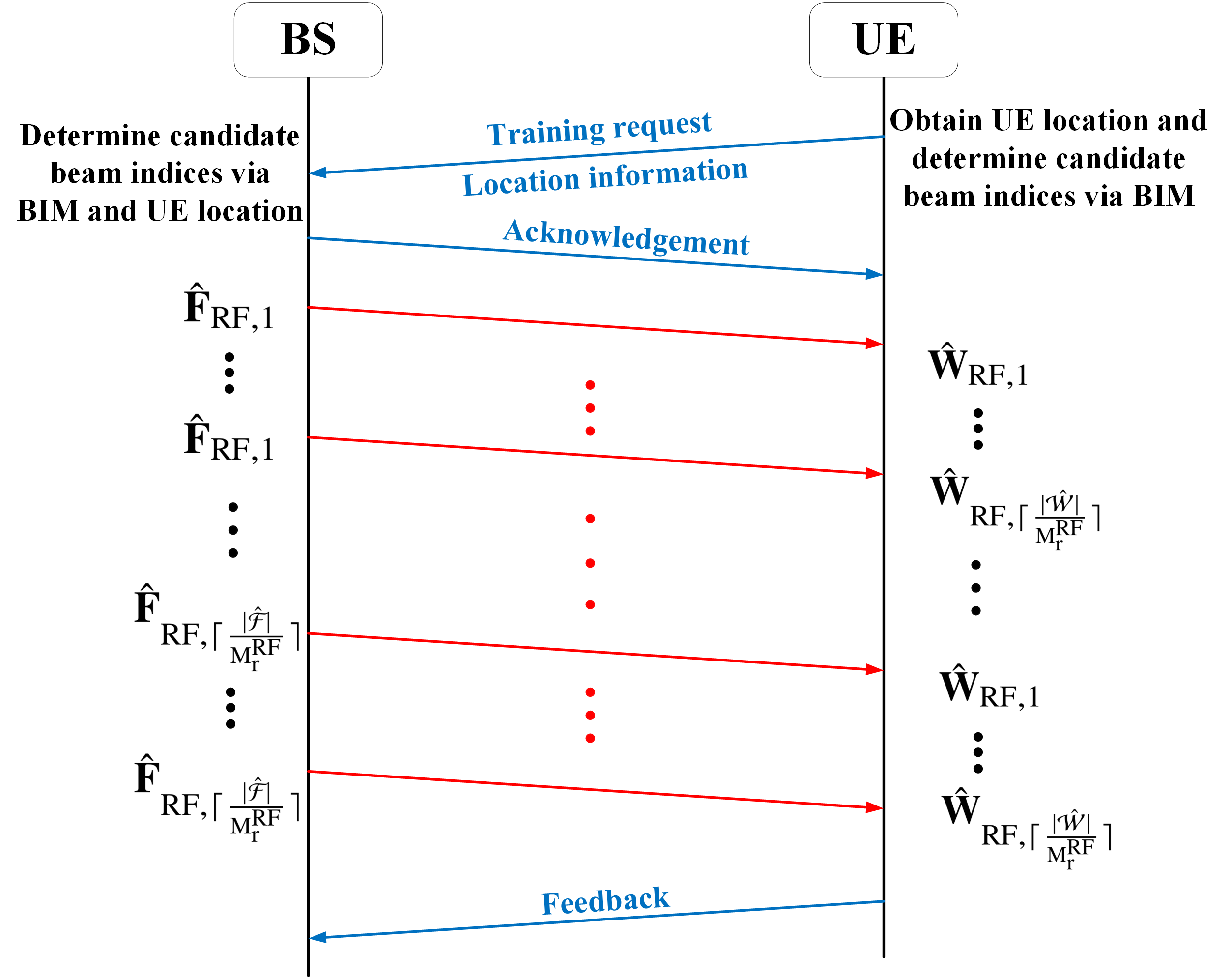}} 
	\caption{Beam sweeping scheme for hybrid beamforming, where the digital beamforming matrices $ \mathbf{F}_\mathrm{BB} $ and $ \mathbf{W}_\mathrm{BB} $ are set as (\ref{FBBWBB}).}
	\label{trainingprocess}
\end{figure}

Similar to (\ref{Ytraining}), each beam sweeping epoch includes $ M_s  $ symbol durations, and let $ \mathbf S \in \mathbb{C}^{M_s \times M_s} $ denote the pilot symbols for the beam sweeping epoch, where $ \mathbf{S}\mathbf{S}^H=\mathbf{I}_{M_s} $.
Thus, when the beam sweeping matrices $ (\hat{\mathbf{F}}_\mathrm{RF,i},\hat{\mathbf{W}}_\mathrm{RF,j}) $ are used, the received signal matrix $ \mathbf{Y}_{ij}\in \mathbb{C}^{M_r^\mathrm{RF}\times M_r^\mathrm{RF}} $ by concatenating the signals over $ M_s $ symbol durations can be written as 
\begin{equation}
	\begin{aligned}
		\mathbf{Y}_{ij}
		=&\sqrt{P}\rho\hat{\mathbf{W}}_\mathrm{RF,j}^H\mathbf{H}\hat{\mathbf{F}}_\mathrm{RF,i}\mathbf{S}+\hat{\mathbf{W}}_\mathrm{RF,j}^H\mathbf{N}_{ij},
	\end{aligned}
\end{equation}
where $ \mathbf{N}_{ij} $ denotes the noise matrix.
After projecting $ \mathbf{Y}_{ij} $ to $ \mathbf{S}^H $, we have
\begin{equation}
	\begin{aligned}
		\tilde{\mathbf{Y}}_{ij}=\mathbf{Y}_{ij}\mathbf{S}^H=\sqrt{P}\rho\hat{\mathbf{W}}_\mathrm{RF,j}^H\mathbf{H}\hat{\mathbf{F}}_\mathrm{RF,i}+\hat{\mathbf{W}}_\mathrm{RF,j}^H\mathbf{N}_{ij}\mathbf{S}^H.
	\end{aligned}
\end{equation}
Note that the $ (p,q) $th element of  $\tilde{\mathbf{Y}}_{ij}\in\mathbb{C}^{M_r^\mathrm{RF}\times M_r^\mathrm{RF}} $ is
\begin{equation}
	\begin{aligned}
		[\tilde{\mathbf{Y}}_{ij}]_{pq}=\sqrt{P}\rho\mathbf{w}_q^H\mathbf{H}\mathbf{f}_p+n_{pq},\
		p=(i-1)M_r^\mathrm{RF}+1,\ldots,iM_r^\mathrm{RF},q=(j-1)M_r^\mathrm{RF}+1,\ldots,jM_r^\mathrm{RF},\\
	\end{aligned}
\end{equation}
where $ n_{pq} $ denotes the corresponding noise element in $ \hat{\mathbf{W}}_\mathrm{RF,j}^H\mathbf{N}_{ij}\mathbf{S}^H $.
Then, after beam sweeping over all the $\lceil\frac{ |\hat{\mathcal{F}}|}{M_r^\mathrm{RF}}\rceil\lceil\frac{ |\hat{\mathcal{W}}|}{M_r^\mathrm{RF}}\rceil$ pairs of beamforming matrices $ (\hat{\mathbf{F}}_\mathrm{RF,i},\hat{\mathbf{W}}_\mathrm{RF,j})$ in (\ref{hatFW}), as illustrated in Fig. \ref{trainingprocess}, which requires $M_r^\mathrm{RF}\lceil\frac{ |\hat{\mathcal{F}}|}{M_r^\mathrm{RF}}\rceil\lceil\frac{ |\hat{\mathcal{W}}|}{M_r^\mathrm{RF}}\rceil$ training symbol durations, we obtain a concatenated matrix $\tilde{\mathbf{Y}}\in \mathbb{C}^{ |\hat{\mathcal{W}}|\times |\hat{\mathcal{F}}|}$ as
\begin{equation}
	\begin{aligned}
	\tilde{\mathbf{Y}}
	=\left[\begin{array}{ccc}
		\tilde{	\mathbf{Y}}_{11} & \cdots &\tilde{	\mathbf{Y}}_{1\lceil\frac{ |\hat{\mathcal{F}}|}{M_r^\mathrm{RF}}\rceil} \\
		\vdots & \ddots & \vdots \\
	\tilde{	\mathbf{Y}}_{\lceil\frac{ |\hat{\mathcal{W}}|}{M_r^\mathrm{RF}}\rceil 1} & \cdots & \tilde{	\mathbf{Y}}_{\lceil\frac{ |\hat{\mathcal{W}}|}{M_r^\mathrm{RF}}\rceil\lceil\frac{ |\hat{\mathcal{F}}|}{M_r^\mathrm{RF}}\rceil}
	\end{array}\right]
		=\left[\begin{array}{ccc}
			y_{11} & \cdots & y_{1 |\hat{\mathcal{F}}|}\\
			\vdots & \ddots & \vdots \\
			y_{|\hat{\mathcal{W}}| 1}& \cdots & y_{|\hat{\mathcal{W}}||\hat{\mathcal{F}}|}
		\end{array}\right].
	\end{aligned}
\label{Ytilde}
\end{equation}

It is observed that $\tilde{\mathbf{Y}}$ in (\ref{Ytilde}) contains the measurement results of all the possible beam pairs $ (\mathbf{f}_p,\mathbf{w}_q) $ in $ \hat{\mathcal{F}} $ and $ \hat{\mathcal{W}} $.
As such, when the training SNR is sufficiently large so that the noise effect can be ignored, the beam selection problem in (\ref{maxFW}) is equivalent to finding a submatrix $ \hat{\mathbf{Y}}\in\mathbb{C}^{M_r^\mathrm{RF}\times M_t^\mathrm{RF}} $ from $\tilde{\mathbf{Y}}$ with the maximum Frobenius norm, i.e., 
\begin{equation}
	\begin{aligned}
		\left(\hat{\mathbf{F}}_\mathrm{RF}, \hat{\mathbf{W}}_\mathrm{RF}\right)
		= \mathop{\arg\max}_{\substack{p_1,\ldots, p_{M_t^\mathrm{RF}} \in \{1,2,\ldots,|\hat{\mathcal{F}}|\}\\q_1,\ldots, q_{M_r^\mathrm{RF}} \in \{1,2,\ldots,|\hat{\mathcal{W}}|\}}} \sum_{m=1}^{M_t^\mathrm{RF}}\sum_{n=1}^{M_r^\mathrm{RF}}  |y_{q_np_m}|^2
		= \arg \max ||\hat{\mathbf{Y}}||_F^2.
		\label{FRFWRF}
	\end{aligned}
\end{equation}
The submatrix searching problem in (\ref{FRFWRF}) can be optimally solved via exhaustive search, which, however, requires prohibitive complexity.
To further reduce the computational complexity, we propose a greedy algorithm, which selects first $ M_t^\mathrm{RF} $ columns with the maximum norms, and then $ M_r^\mathrm{RF} $ rows. 
The main procedures are summarized in Algorithm 2.

After obtaining $ \hat{\mathbf{Y}} $ and  the corresponding analog beamforming matrices $ \hat{\mathbf{F}}_\mathrm{RF} $ and $ \hat{\mathbf{W}}_\mathrm{RF} $ via Algorithm 2, the digital beamforming matrices can be easily obtained based on (\ref{Basebandbeamforming}) by setting the equivalent channel in (\ref{Htilde}) as
\begin{equation}
	\begin{aligned}
		\tilde{\mathbf{H}}=\frac{1}{\rho\sqrt{P}}(\hat{\mathbf{W}}_\mathrm{RF}^H\hat{\mathbf{W}}_\mathrm{RF})^{-\frac{1}{2}}\hat{\mathbf{Y}}(\hat{\mathbf{F}}_\mathrm{RF}^H\hat{\mathbf{F}}_\mathrm{RF})^{-\frac{1}{2}}.
	\end{aligned}
\end{equation}

\begin{algorithm}[t]
	\caption{Finding submatrix with maximum Frobenius norm} 
	\hspace*{0.02in} {\bf Input:} 
	The measurement matrix $\tilde{\mathbf{Y}}\in \mathbb{C}^{|\hat{\mathcal{W}}|\times|\hat{\mathcal{F}}|}$\\
	\hspace*{0.02in} {\bf Output:} 
	The submatrix $\hat{ \mathbf{Y}}\in \mathbb{C}^{M_t^\mathrm{RF}\times M_r^\mathrm{RF}}$, and the analog beamforming matrices $ \hat{\mathbf{F}}_\mathrm{RF}$ and $  \hat{\mathbf{W}}_\mathrm{RF}$
	\begin{algorithmic}[1]
		\State Initialize $ \mathcal{R}_1=\emptyset,\mathcal{R}_2=\emptyset $
		\For{$ m=1,\ldots,M_t^\mathrm{RF}$} 
		\State $ p_m=\arg \max_{i\notin \mathcal{R}_1}||[\tilde{\mathbf{Y}}]_{:,p}||_2$
		\State $ \mathcal{R}_1=\mathcal{R}_1\cup\{p_m\} $
		\EndFor
		\For{$ n=1,\ldots,M_r^\mathrm{RF}$} 
		\State $q_n=\arg \max_{q\notin \mathcal{R}_2}||[\tilde{\mathbf{Y}}]_{j,\mathcal{R}_1}||_2$
		\State $ \mathcal{R}_2=\mathcal{R}_2\cup\{q_n\} $
		\EndFor
		\State $\hat{ \mathbf{Y}}=[\tilde{\mathbf{Y}}]_{\mathcal{R}_2,\mathcal{R}_1} $
		\Statex $ \hat{\mathbf{F}}_\mathrm{RF}= [\mathbf{f}_{p_1},\mathbf{f}_{p_2},\ldots,\mathbf{f}_{p_{M^\mathrm{RF}_t}}]$
		\Statex $ \hat{\mathbf{W}}_\mathrm{RF}= [\mathbf{w}_{q_1},\mathbf{w}_{q_2},\ldots,\mathbf{w}_{q_{M^\mathrm{RF}_r}}] $
		\State \Return $\hat{ \mathbf{Y}},\hat{\mathbf{F}}_\mathrm{RF},\hat{\mathbf{W}}_\mathrm{RF}$
	\end{algorithmic}
\end{algorithm}
Note that different from the beam sweeping schemes over the complete codebooks $ \mathcal{F}$ and $\mathcal{W} $, the BIM-enabled beam sweeping can significantly reduce the number of required training beam pairs from $  |\mathcal{F}| |\mathcal{W}| $ to $  |\hat{\mathcal{F}}| |\hat{\mathcal{W}}|  $, since we usually have $M^\mathrm{RF}_t \leq |\hat{\mathcal{F}}| \ll  M_t \leq |\mathcal{F}| $ and $M^\mathrm{RF}_r \leq |\hat{\mathcal{W}}| \ll M_r \leq  |\mathcal{W}| $. 
Thus, the effective communication rate becomes 
\begin{equation}
	\begin{aligned}
		R=&\frac{1}{T} \sum_{t=1}^{T} \frac{N-N_{t r}}{N}\log _{2} \left|\mathbf{I}_{M_r^\mathrm{RF}}+\tilde{P}\hat{\mathbf{H}}_\mathrm{e}[t]\hat{\mathbf{R}}_x[t]\hat{\mathbf{H}}_\mathrm{e}^H[t] \right|\\
		\leq&\left(1-\frac{N_{t r}}{N}\right)R_\mathrm{opt}\\
		=&\left(1-\frac{M_r^\mathrm{RF}}{N}\lceil\frac{ |\hat{\mathcal{F}}[t]|}{M_r^\mathrm{RF}}\rceil\lceil\frac{ |\hat{\mathcal{W}}[t]|}{M_r^\mathrm{RF}}\rceil\right) R_\mathrm{opt},
		\label{BIMrate}
	\end{aligned}
\end{equation}
where the first inequality is due to the fact that the digital beamforming matrices is determined based on the resulting equivalent channel after analog beam selection;
and the second equality follows since the proposed beam sweeping requires $N_\mathrm{tr}= M_r^\mathrm{RF}\lceil\frac{ |\hat{\mathcal{F}}[t]|}{M_r^\mathrm{RF}}\rceil\lceil\frac{ |\hat{\mathcal{W}}[t]|}{M_r^\mathrm{RF}}\rceil$ training symbol durations.
By comparing (\ref{trainrate}) and (\ref{BIMrate}), it is observed that the BIM-enabled hybrid beamforming can significantly reduce the pre-log factor in (\ref{trainrate}) and thus approach $ R_\mathrm{opt} $ more closely.
Besides, the BIM-enabled scheme can significantly reduce the computational complexity, since only one matrix SVD is computed once.

It is observed from Fig. \ref{CAM3} and Fig. \ref{BIM3} that compared with CAM-enabled hybrid beamforming, the BIM-based counterpart not only simplifies the implementation, but also reduces the required storage space (since finite number of beam indices rather than continuous path information needs to be obtained) and the computational complexity (with only one matrix SVD required).
On the other hand, compared with BIM, the main advantages of CAM lie in that it provides the information of the intrinsic radio propagation channel characteristics, and is independent of the array configuration or the beamforming codebook.
Thus, the choice between CAM and BIM depends on the practical application scenarios.  

\subsection{Construction of CAM and BIM}
The performance of the proposed CAM/BIM-enable beam alignment scheme depends on the accuracy of the UE location and the CAM/BIM construction.
With the development of localization technologies, the acquisition of sufficiently accurate UE location becomes more feasible than ever before.
For example, many localization technologies such as light detection and ranging (LiDAR) and real-time kinematic (RTK) have been widely used to achieve centimeter level localization.
As discussed in \cite{wu2021environment} and \cite{CKM}, the CAM/BIM depends on the data availability, which could be obtained via the offline/online data collection.
The candidate sets can be acquired via an extra step to store the channel training results for online data acquisition.
Another way is to use the offline schemes. 
With the help of the simulation method such as ray tracing, the site-specific channel can be reconstructed by the obtained path information using the available physical environment information (such as 3D map), based on which the candidate channel knowledge sets can be obtained.
Some special measuring equipment can also be used to obtain the site-specific CSI.
The above online/offline methods can be used in combination to get sufficient data.
When finite data points have been collected, various machine learning techniques such as K-nearest neighbor (KNN), Kridging-based method, or Deep Neural Networks (DNNs) could be applied to learn the entire CKM.
Besides, in \cite{li2021channel}, an EM algorithm-based method is proposed for CKM construction.
Note that the cost of the CKM construction is independent of the real-time communication and will not occupy the communication time.

\section{Simulation Results}
\begin{figure}[htbp]
	\vspace{-0.8cm}
	\centering{\includegraphics[width=.45\textwidth]{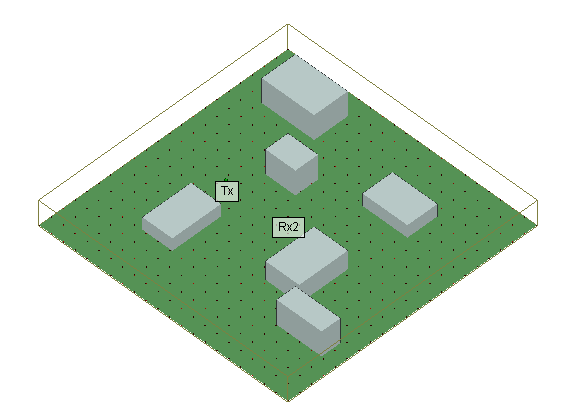}} 
	\caption{Physical environment for numerical simulations.} 
	\label{Scene}  
\end{figure}

We consider a physical environment shown in Fig. \ref{Scene}, where one BS with location labelled as `Tx' is shown. The commercial ray tracing software Remcom Wireless Insite is used to generate the ground-truth channel information, including the power, phase, AoA, AoD of the channel paths at each location. 
One example receiver location is also shown in the figure. 

The BS is assumed to be equipped with $ M_t^\mathrm{RF}=4 $ RF chains and $ M_t^z \times M_t^y $ uniform planar array (UPA) with adjacent elements separated by half wavelength, where $ M_t^z $ and $ M_t^y $ varies from 8 to 20. 
Therefore, the total number of BS antennas $ M_t $ varies from 64 to 400. The Kronecker product based beamforming codebook \cite{codebook} is used.
Furthermore, the UE is assumed to be equipped with $ M_r^\mathrm{RF}=4 $ RF chains and $4\times4$ UPA, with a similar Kronecker product based beamforming codebook. 
Therefore, the maximum number of data streams that can be supported is $ M_s=4$.

A CAM is constructed to learn a maximum of $ \hat{L}=40 $ candidate channel paths for each UE location, which gives their corresponding zenith and azimuth AoDs and AoAs. 
Based on the generated channels for 3700 UE locations, the inverse distance weighting (IDW) method of the K-nearest neighbors (KNN, with $ K=3 $) is used to construct the entire CAM of all locations.
Furthermore, a BIM is constructed to learn 20 candidate beams for the BS and 10 candidate beams for the UE at each location. 
The training data are obtained with the selected beams for 3700 randomly selected UE locations. 
The KNN method with $ K=3 $ is used to learn the entire BIM based on such finite training samples.

For the benchmark schemes, we consider the conventional LS-based channel estimation and the channel estimation method based on OMP in \cite{lee2016channel}.
Besides, we also consider another technique utilizing UE location, namely the location-based beam alignment \cite{location}, where the AoAs/AoDs are directly calculated based on the relative positions of the BS and UE, but ignores the actual propagation environment.
To obtain the performance upper bound, the optimal solution in Algorithm 1 in Section III is applied based on perfect CSI. 
The channel coherent time is assumed to span over $ N=1200 $ symbols, and all simulations given below are averaged over 300 randomly selected UE locations. 

\begin{figure}[htbp]
	\vspace{-1cm}
	\setlength{\abovecaptionskip}{-0.2cm}
	\centering
	\subfigure[]{
			\begin{minipage}[t]{0.5\textwidth}
					\centering
					\includegraphics[width=1\textwidth]{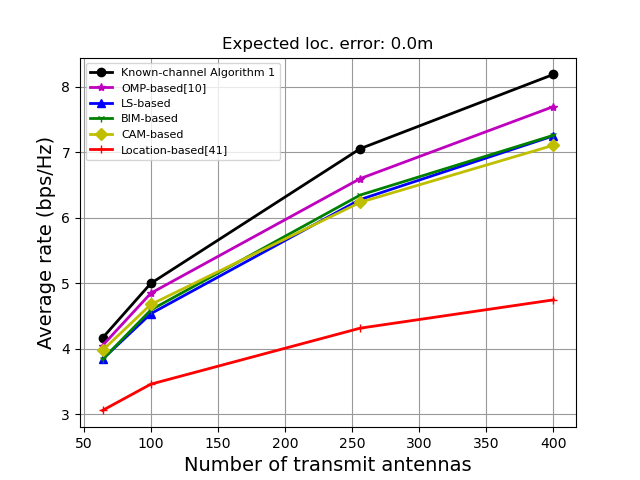}
				\end{minipage}%
			\label{Rate0}
		}%
	\subfigure[]{
			\begin{minipage}[t]{0.5\textwidth}
					\centering
					\includegraphics[width=1\textwidth]{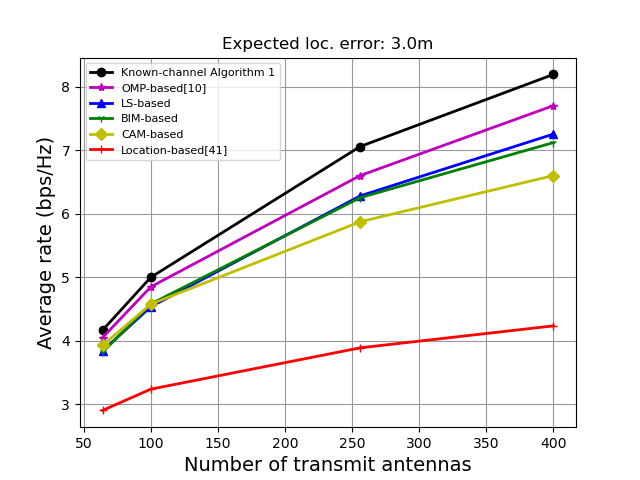}
				\end{minipage}%
			\label{Rate3}
		}%
	\centering
	\caption{Comparison of average effective communication rate for various hybrid beamforming schemes (without considering the impact of training overhead). (a) The UE location is assumed to be perfectly known. (b) The UE location is assumed to have an average error of 3 meters.}
	\label{Rate}
\end{figure}

Fig. \ref{Rate0} shows the average communication rate versus the number of transmit antennas $ M_t $ for various hybrid beamforming schemes, without considering the impact of the training overhead for the time being.
For the location-based and CKM-based schemes, the UE location is assumed to be perfectly known. 
As shown in Fig. \ref{Rate0}, without considering the training overhead, the communication rate of the OMP-based channel estimation scheme in \cite{lee2016channel} is slightly better than the proposed CAM- and BIM-enabled hybrid beamforming schemes, due to its extensive training applied.
Furthermore, it is observed that the location-based scheme results in poor performance for all $ M_t $ considered, since it is ignorant of the actual propagation environment. 
In contrast, the proposed CAM- and BIM-enabled hybrid beamforming schemes significantly outperform the location-based scheme, thanks to its environment-awareness enabled by CKM. 
Besides, both proposed schemes yield performance close to the upper bound that is obtained based on perfect CSI, e.g., around 90\% when the number of transmit antennas is 400. 
It is further observed from Fig. \ref{Rate0} that the BIM-based scheme achieves slightly better performance than CAM-based scheme for moderate $ M_t $, due to its end-to-end overall learning capability, without relying on the intermediate channel reconstruction. 

To study the impact of UE location (loc.) error on the performance of the proposed schemes, Fig. \ref{Rate3} plots the average communication rate for various schemes with average location error of 3 meters, which is modeled based on Rayleigh distribution. 
By comparing Fig. \ref{Rate0} and Fig. \ref{Rate3}, it is observed that the location error only slightly degrades the rate performance of the proposed CAM- and BIM-enabled schemes. Furthermore, compared to CAM, BIM is less affected by location error, since the beams are expected to be less sensitive to location variations than angle information.

%

Fig. \ref{Training_overhead} shows the required number of training intervals for various hybrid beamforming schemes.
Compared with conventional LS-based channel estimation scheme, the proposed CAM- and BIM-enabled schemes can drastically reduce the number of training time slots, e.g., from more than 1000 time slots to less than 50 time slots for $ M_t=256 $.
Compared with the OMP-based scheme, the proposed CAM- and BIM-enabled schemes also achieve a significant reduction on the required training time slots.
The analytical comparison of training overhead among different schemes is shown in Table I.

\begin{table}[h] 
\vspace{1.0em}
\setlength{\abovecaptionskip}{0cm}
\centering
\renewcommand{\arraystretch}{1.3}
\caption{Comparison of training overhead}
\begin{tabular}{p{1.5cm}p{1.5cm}p{1.8cm}p{2.2cm}p{1.9cm}}
	\toprule[1.3pt]
	{\textbf{Methods}} &  LS-based  & CAM-based & BIM-based & OMP-based\\[2.0pt]
	\hline \textbf{Overhead} & $ \lceil \frac{M_tM_r}{M_r^\mathrm{RF}}\rceil$ &$\lceil\frac{\hat{L}[t]}{M_r^\mathrm{RF}}\rceil $&$\lceil\frac{ |\hat{\mathcal{F}}[t]| |\hat{\mathcal{W}}[t]|}{M_r^\mathrm{RF}}\rceil $&$\mathcal{O}(\lceil\frac{L[t]\ln( |\bar{\Omega}|)}{M_r^\mathrm{RF}}\rceil) $\\[4pt]
	\bottomrule[1.3pt]
\end{tabular}
\vspace{-2.0em}
\end{table}

\begin{figure}[htbp]
	\vspace{-1cm}
	\setlength{\abovecaptionskip}{-0.05cm}
	\setlength{\belowcaptionskip}{-1.1cm}
	\centering{\includegraphics[width=.5\textwidth]{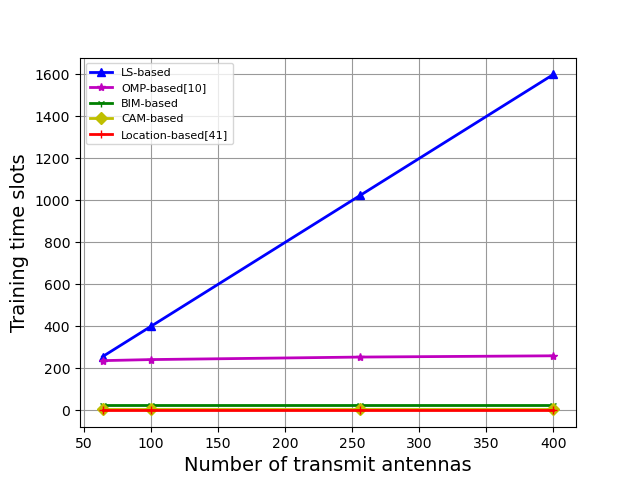}}  	 		 		\caption{Comparison of training overhead for various hybrid beamforming schemes.} \label{Training_overhead}  	 		 	 		 	
\end{figure}

\begin{figure}[htbp]
	\vspace{-0.79cm}
	\setlength{\abovecaptionskip}{-0.2cm}
	\centering
	\subfigure[]{
		\begin{minipage}[t]{0.5\textwidth}
			\centering
			\includegraphics[width=1\textwidth]{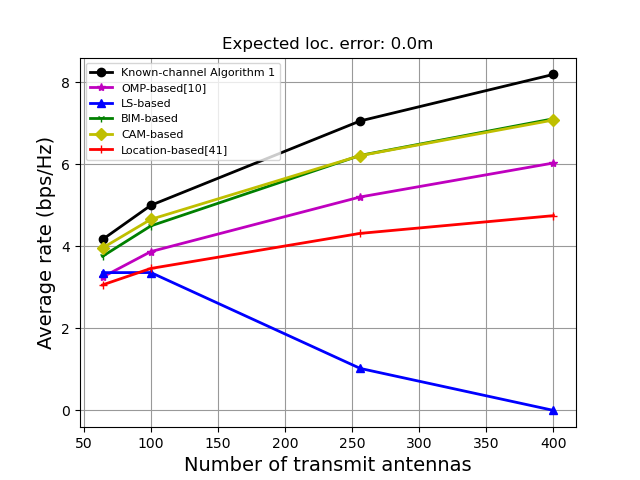}
		\end{minipage}%
		\label{True_rate0}
	}%
	\subfigure[]{
		\begin{minipage}[t]{0.5\textwidth}
			\centering
			\includegraphics[width=1\textwidth]{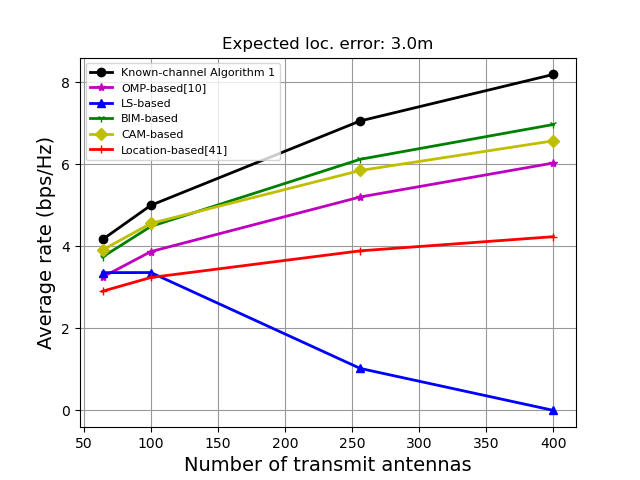}
		\end{minipage}%
		\label{True_rate3}
	}%
	\centering
	\caption{Comparison of average effective communication rate for various hybrid beamforming schemes. (a) The UE location is assumed to be perfectly known. (b) The UE location is assumed to have an average error of 3 meters.}
	\label{TrueRate}
\end{figure}

With the impact of training overhead taken into account, the average effective communication rate is shown in Fig. \ref{True_rate0}. 
It is observed that the effective communication rate of the LS-based channel estimation scheme decreases drastically as the number of BS antennas $ M_t $ grows, since its training overhead outweighs the resulting beamforming gain for large antenna systems.
By contrast, the proposed CAM- and BIM-enabled schemes give monotonically increasing performance as $ M_t $ increases, and they both significantly outperform the OMP-based channel estimation scheme and the location-based scheme, thanks to its drastic reduction of the training overhead with environment awareness.

%

Fig. \ref{True_rate3}  plots the average effective communication rate for various schemes with average location error of 3 meters, which is modeled based on Rayleigh distribution.
It is shown that even with such a moderate location error, the two proposed schemes still significantly outperform the various benchmark hybrid beamforming schemes.
This is expected since the moderate UE location error slightly degrades the resulting beamforming gain that only affects the rate logarithmically, while the saving of the training overhead that affects the communication rate linearly is more dominating. This demonstrates the great potential of the proposed CKM-enabled hybrid beamforming for large-scale MIMO systems.

\section{Conclusion}
In this paper, a novel environment-aware hybrid beamforming technique enabled by CKM is proposed, which can greatly reduce the prohibitive training overhead for the mmWave massive MIMO systems, while achieving high rate performance. 
Two specific types of CKM are introduced, namely CAM and BIM, and their enabled hybrid beamforming schemes are presented in detail. 
Simulation results over practical propagation environment based on ray-tracing demonstrate that the proposed schemes significantly outperform the benchmark schemes, even with moderate errors on the UE location.



%

\appendix[Proof of Lemma1]
If $ \mathrm{vec}(\tilde{\mathbf{N}})\sim \mathcal{CN}(\mathbf{0},\sigma^2\mathbf{I}_{M_s^2}) $, the MSE in (\ref{MSE}) can be written as 
\begin{align}
		\mathrm{MSE}
		=&\mathbb{E}\left[\left| \left| \frac{1}{\sqrt{P}}(\hat{\mathbf{Q}}^H\hat{\mathbf{Q}})^{-1}\hat{\mathbf{Q}}^H\mathrm{vec}(\tilde{\mathbf{N}})\right| \right| _2^2\right]\notag\\
		=&\frac{\sigma^2}{P}\mathrm{tr}((\hat{\mathbf{Q}}^H\hat{\mathbf{Q}})^{-1})\notag\\
		=&\frac{\sigma^2}{P}\sum_{l=1}^{\hat{L}}\lambda_l((\hat{\mathbf{Q}}^H\hat{\mathbf{Q}})^{-1})\notag\\
		=&\frac{\sigma^2\hat{L}}{P}\left(\frac{1}{\hat{L}}\sum_{l=1}^{\hat{L}}\lambda_l((\hat{\mathbf{Q}}^H\hat{\mathbf{Q}})^{-1})\right)\notag\\
		\overset{(a)}{\geq}&\frac{\sigma^2\hat{L}^2}{P\sum_{l=1}^{\hat{L}}1/\lambda_l((\hat{\mathbf{Q}}^H\hat{\mathbf{Q}})^{-1})}\notag\\
		\overset{(b)}{=}&\frac{\sigma^2\hat{L}^2}{P\sum_{l=1}^{\hat{L}}\lambda_l(\hat{\mathbf{Q}}^H\hat{\mathbf{Q}})}\notag\\
		=&\frac{\sigma^2\hat{L}^2}{P\mathrm{tr}(\hat{\mathbf{Q}}^H\hat{\mathbf{Q}})}\notag\\
		\overset{(c)}{=}&\frac{\sigma^2\hat{L}^2}{\parbox{5.3cm}{$ P\mathrm{tr}((\mathbf{A}_t^T(\hat{\Phi})\mathbf{F}_\mathrm{RF}^*\mathbf{F}_\mathrm{BB}^*\mathbf{F}_\mathrm{BB}^T\mathbf{F}_\mathrm{RF}^T\mathbf{A}_t^*(\hat{\Phi}))*(\mathbf{A}_r^H(\hat{\Theta})\mathbf{W}_\mathrm{RF}\mathbf{W}_\mathrm{BB}\mathbf{W}_\mathrm{BB}^H\mathbf{W}_\mathrm{RF}^H\mathbf{A}_r(\hat{\Theta}))) $}}\notag\\
		\overset{(d)}{\geq}&\frac{\sigma^2\hat{L}^2}{\parbox{5.3cm}{$ P\mathrm{tr}(\mathbf{A}_t^H(\hat{\Phi})\mathbf{F}_\mathrm{RF}\mathbf{F}_\mathrm{BB}\mathbf{F}_\mathrm{BB}^H\mathbf{F}_\mathrm{RF}^H\mathbf{A}_t(\hat{\Phi}))\times\mathrm{tr}(\mathbf{A}_r^H(\hat{\Theta})\mathbf{W}_\mathrm{RF}\mathbf{W}_\mathrm{BB}\mathbf{W}_\mathrm{BB}^H\mathbf{W}_\mathrm{RF}^H\mathbf{A}_r(\hat{\Theta})) $}}\notag\\
		=&\frac{\sigma^2\hat{L}^2}{P||\mathbf{A}_t^H(\hat{\Phi})\mathbf{F}_\mathrm{RF}\mathbf{F}_\mathrm{BB}||_2^F||\mathbf{A}_r^H(\hat{\Theta})\mathbf{W}_\mathrm{RF}\mathbf{W}_\mathrm{BB}||_2^F}
		\label{MSE2}
\end{align}
where $ (a) $ follows because the arithmetic mean is larger than the harmonic mean\cite{bullen2013handbook}, $ (b) $ holds because $1/\lambda_l((\hat{\mathbf{Q}}^H\hat{\mathbf{Q}})^{-1})=\lambda_l(\hat{\mathbf{Q}}^H\hat{\mathbf{Q}})$, $ (c) $ follows from the property of Khatri-Rao product, i.e., $ (\mathbf{A}\circ\mathbf{B})^H(\mathbf{A}\circ\mathbf{B})=\mathbf{A}^H\mathbf{A}* \mathbf{B}^H\mathbf{B}$\cite{liu2008hadamard}, and $ (d) $ is due to the Cauchy-Schwarz inequality.


%

%

%
%

\ifCLASSOPTIONcaptionsoff
  \newpage
\fi



\bibliographystyle{IEEEtran}
\bibliography{ref2}

\end{document}